%% file: main.tex
\documentclass{article}

\usepackage[final,nonatbib]{neurips_2022}

\usepackage[utf8]{inputenc} 
\usepackage[T1]{fontenc}    
\usepackage{hyperref}       
\usepackage{url}            
\usepackage{booktabs}       
\usepackage{amsfonts}       
\usepackage{nicefrac}       
\usepackage{microtype}      
\usepackage{xcolor}         
\usepackage{style}
\title{Diversified Recommendations for Agents with Adaptive Preferences}

\author{%
  Arpit Agarwal\thanks{http://www.columbia.edu/~aa4931/} \\
  Department of Computer Science\\
  Columbia University\\
  New York, NY 10027 \\
  \texttt{arpit.agarwal@columbia.edu} \\
\And
  William Brown\thanks{wibrown.github.io} \\
  Department of Computer Science\\
  Columbia University\\
  New York, NY 10027 \\
  \texttt{w.brown@columbia.edu} \\
}

\begin{document}

\maketitle

\begin{abstract}

When an Agent visits a platform recommending a menu of content to select from, their choice of item depends not only on immutable preferences, but also on their prior engagements with the platform. The Recommender's primary objective is typically to encourage content consumption which optimizes some reward, such as ad revenue, but they often additionally aim to ensure that a sufficiently wide variety of content is consumed by the Agent over time. We formalize this problem as an adversarial bandit task. At each step, the Recommender presents a menu of $k$ (out of $n$) items to the Agent, who selects one item in the menu according to their unknown {\it preference model}, which maps their history of past items to relative selection probabilities. The Recommender then observes the Agent's selected item and receives bandit feedback of the item's (adversarial) reward. In addition to optimizing  reward from the selected items at each step, the Recommender must also ensure that the total distribution of chosen items has sufficiently high entropy. 

We define a class of preference models which are {\it locally learnable}, i.e.\ behavior over the entire domain can be estimated by only observing behavior in a small region; this includes models representable by bounded-degree polynomials as well as functions with a sparse Fourier basis. For this class, we give an algorithm for the Recommender which obtains $\Tilde{O}(T^{3/4})$ regret against all item distributions satisfying two conditions: they are sufficiently diversified, and they are {\it instantaneously  realizable} at any history by some distribution over menus. We show that these conditions are closely connected:  all sufficiently high-entropy distributions are instantaneously realizable at any history of selected items. We also give a set of negative results justifying our assumptions, in the form of a runtime lower bound for non-local learning and linear regret lower bounds for alternate benchmarks.

\end{abstract}

\input{intro}

\input{prelims}

\input{contracting}

\input{uniform}

\input{future}

\clearpage
\bibliography{ref}


\clearpage
\appendix

\input{appendix}

\end{document}

%% file: intro.tex
\section{Introduction}\label{sec:intro}

Suppose you manage an online platform that repeatedly provides menus of recommended content to visitors, such as sets of videos to watch or items to purchase, 
aiming to display options which agents will engage favorably with and yield you high rewards (in the form of ad revenue, watch time, purchases, or other metrics). In many settings, the preferences of agents are not fixed {\it a priori}, but rather can \emph{change} as a function of their consumption patterns---the deeper one goes down a content ``rabbit hole'', the further one might be likely to keep going. 
This ``rabbit hole'' effect can lead to (unforeseen) loss of revenue for the platform, as advertisers may later decide that they are not
willing to pay as much for this ``rabbit hole'' content as they would for other content.
The scope of negative effects emerging from these feedback loops is large,
ranging from 
the emergence of ``echo chambers'' \cite{DBLP:journals/corr/abs-2007-02474}
and rapid political polarization \cite{o2015down} to 
increased homogeneity which can decrease agent utility \cite{DBLP:journals/corr/abs-1710-11214}, amplify bias \cite{DBLP:journals/corr/abs-2007-13019}, or drive content providers to leave the platform \cite{DBLP:journals/corr/abs-2008-00104}. These are harms which many platforms aim to avoid, both for their own sake and out of broader societal concerns. 

Hence, the evolving preferences of the Agent can be directly 
at odds with 
the Recommender's objectives of maximizing revenue and ensuring diverse consumption patterns in this \emph{dynamic} environment.
Our goal is to study such tensions between 
the interaction of these two players:
the Recommender that recommends menus based on past choices of the Agent so as to maximize its reward (subject to 
diversity constraints), and 
the Agent whose preferences evolve as a function of past recommendations.

To this end, we consider a stylized setting where the Recommender is tasked with providing a {\it menu} of $k$ recommended items (out of $n$ total) every round to an Agent for $T$ sequential rounds. 
In each round, the Agent observes the menu, then selects one of the items according to their {\it preference model} $M$, which the Recommender does not know in advance. The preference model $M$ takes as input the Agent's {\it memory vector} $v$, which is the normalized histogram of their past chosen items, and assigns relative selection probabilities to each item. 
The selected item at each round results in a reward for the Recommender, specified by an adversarial sequence of reward vectors, 
which the Recommender receives as bandit feedback, in addition to observing which item was selected. The Recommender must choose a sequence of menus to maximize their reward (or minimize regret), subject to a {\it diversity constraint}, expressed as a minimum entropy for the empirical item distribution.

However, any regret minimization problem is incomplete without an appropriate benchmark for comparing the performance 
of a learner. An entropy constraint alone is insufficient to define a such benchmark. Due to intricacies of the Agent's preference model, there may be item distributions which are impossible to induce under any sequence of menus (e.g.\ they may strongly dislike the most profitable content).
Adding to the challenge is the fact that the preference model is initially unknown and must be {\it learned}, and the set of item distributions which are {\it instantaneously realizable} by sampling a menu from some distribution can shift each round as well. Several immediate proposals are infeasible: it is impossible to obtain sublinear regret against the best fixed menu distribution, or even against the best item distribution  realizable from the uniform memory vector. 
We propose a natural benchmark for which regret minimization becomes possible: the set of item distributions which are {\it everywhere}  instantaneously realizable (the $\texttt{EIRD}(M)$ set), i.e.\ item distributions such that, at any memory vector, there is always {\it some} menu distribution which induces them. We show that this set is also closely related to entropy constraints: when $M$ is sufficiently {\it dispersed} (a condition on the minimum selection probability for each item), $\texttt{EIRD}(M)$ contains all sufficiently high-entropy distributions, and so regret minimization can  occur over the entire high-entropy set.

\subsection{Our Results}

We give an algorithm which, for a minimum entropy set $H_c$ and preference model $M$, allows the Recommender to obtain $\Tilde{O}(T^{3/4 })$ regret 
against the best distribution in the intersection of $H_c$ and $\texttt{EIRD}(M)$, provided that $M$
satisfies $\lambda$-{\it dispersion} 
and 
belongs to a class $\M$ which is 
{\it locally learnable}.
A $\lambda$-dispersed preference model $M$ assigns a preference score of at least $\lambda > 0$ to every item, ensuring a minimum positive probability of selection to each item in a menu. Dispersion is a natural assumption, given our restriction to $\texttt{EIRD}(M)$, as items
which only have positive selection probability in part of the domain cannot be induced everywhere. 
The local learnability condition for a model class enforces that the behavior of any particular model can be predicted by observing behavior only in a small region. This is essentially necessary to have any hope of model estimation in this setting: we show that if learning a class from {\it exact} queries requires making queries to many points which are pairwise well-separated, exponentially many rounds are required to implement query learning. Despite this restriction, we show that several rich classes of preference models are indeed locally learnable, including those where preference scoring functions are expressed by bounded-degree multivariate polynomials, or by univariate functions with a sparse Fourier basis.

Our algorithm is explicitly separated into learning and optimization stages. The sole objective for the learning stage is to solve the {\it outer problem}: recover an accurate hypothesis for the preference model. We select sequences of menus which move the Agent's memory vector to various points near the uniform distribution, enabling us to implement local learning and produce a model hypothesis $\hat{M}$. 
We then shift our focus to the {\it inner problem} for the Recommender, which is natural to view as a bandit linear optimization problem over the set of distributions in consideration, as we can use $\hat{M}$ to identify a distribution of menus which generates a particular item distribution. However, representing the $\texttt{EIRD}(\hat{M})$ set explicitly is impractical, as the functions which generate feasible sets from the history can be highly non-convex. Instead, we operate over the potentially larger set where intersections are taken only over the sets $\texttt{IRD}(v, \hat{M})$ of instantaneously realizable distributions we have observed thus far. This precludes us from using off-the-shelf bandit linear optimization algorithms as a black box, as they typically require the decision set to be specified in advance. We introduce a modification of the \textsc{FKM} algorithm \cite{DBLP:journals/corr/cs-LG-0408007}, \textsc{RC-FKM}, which can operate over contracting decision sets, and additionally can account for the imprecision in $\hat{M}$ when generating menu distributions. This enables the Recommender to guide the Agent to minimize regret on their behalf via the sequence of menus they present.

\subsection{Summary of Contributions}
Briefly, our main contributions are:
\begin{enumerate}
    \item We formulate the dynamic interaction between a Recommender and an Agent as an adversarial bandit task.
    We show that no algorithm can obtain $o(T)$ regret against 
    the best menu distribution, or against the best item distribution in the \texttt{IRD} set of uniform vector.
    We then consider $\texttt{EIRD}(M)$ and argue that it is a natural benchmark for regret as it also contains all sufficiently high entropy distributions over items.
    \item We define a class of \emph{locally learnable}
    functions, which are functions that can be learned only using samples from a small neighborhood. We show a number of rich classes of functions where this is possible, and further we show that any class which is {\it not} locally learnable cannot be learned quickly by any algorithm which fits a hypothesis using queries.
    \item We give an algorithm for the Recommender that achieves $\widetilde{O}(T^{3/4})$ regret against $\texttt{EIRD}(M)$ for locally learnable 
    classes of preference models that are $\lambda$-dispersed, which implements local learning to obtain a sufficiently accurate hypothesis for use in optimizing menu distributions.
     As a component of this, we develop a new algorithm for bandit linear optimization which can operate over contracting decision sets, and which can account for bounded adversarial imprecision in the played action. 
\end{enumerate}

Overall, by considering this stylized setting we are able to provide several insights into the dynamic interaction between an Agent and a Recommender.
While our algorithm is a useful tool for a Recommender who is already committed to providing diversified recommendations, we also view our results as presenting an intrinsic argument for incorporating such constraints.
When preferences adapt over time, and Agents may be prone to venturing down content ``rabbit holes'', restricting attention to recommendation patterns which are not too concentrated on small sets of items
can in fact make the regret minimization problem tractable by discouraging consumption patterns which may be difficult to draw the Agent back from.
This suggests a synergy between the goal of regret minimization 
and showing diverse content to the user.

\subsection{Related Work} 

Feedback loops in user preferences
have received significant attention in the recommender systems literature, particularly for models with multiple agents which make use of collaborative filtering methods, and with explicit adaptivity models which are less flexible than those we consider \cite{DBLP:journals/corr/abs-2201-07203, DBLP:journals/corr/abs-1710-11214, DBLP:journals/corr/abs-2007-13019, 10.5555/3157382.3157461, DBLP:journals/corr/abs-2008-00104}.
Within the online learning literature, our formalization bears some resemblance to bandit problems where multiple arms can be pulled simultaneously, which have received much recent attention \cite{10.1145/1553374.1553527, YUE20121538, DBLP:journals/corr/abs-1807-11398, NEURIPS2020_d5fcc35c}. Our results also share similarities with work on optimization from revealed preferences, where a mapping to a nested convex problem must be learned \cite{DBLP:journals/corr/RothUW15, DBLP:journals/corr/abs-1710-07887}; with the performative prediction literature, where actions induce a distribution shift which impacts instantaneous reward potential \cite{DBLP:journals/corr/abs-2002-06673, DBLP:journals/corr/abs-2202-00628}; and more broadly, with repeated game problems against adaptive agents \cite{DBLP:journals/corr/abs-1711-09176, DBLP:journals/corr/abs-1909-13861, DBLP:journals/corr/DaskalakisS15}. Further related work is discussed in Appendix \ref{sec:app-related}.

\subsection{Organization}
In Section \ref{sec:prelims}, we introduce our setting and key definitions, analyze the local learnability of several classes of preference models, and give a series of negative and structural results.
In Section \ref{sec:contracting} we introduce a bandit linear optimization algorithm for contracting sets, which we use as a subroutine for our main algorithm in Section \ref{sec:uniform}. We discuss the intuition for our proof techniques throughout, with full proofs deferred to the appendix.

%% file: prelims.tex
\section{Model and Preliminaries}
\label{sec:prelims}

The central object of our setting is the {\it preference model} of the Agent, which dictates their relative item preferences based on their selection history and expresses their adaptivity over time. 

\begin{definition}[Preference Models]
A preference model is a mapping $M : \Delta(n) \rightarrow [0,1]^n$ which maps memory vectors $v$ to a preference score vector $s_v = M(v)$. 
\end{definition}

We assume that any input $v \notin \Delta(n)$ to $M$ (such as the empty history at $t=1$) results in the uniform score vector where $M(v)_i = 1$ for all $i$. A constraint on our sequence of interactions with the Agent is that the resulting item distribution must have sufficiently high entropy.

\begin{definition}[Diversity Constraints]
A {\it diversity constraint} $H_c \subset \Delta(n)$ is the convex set containing all item distributions $v \in \Delta(n)$ with entropy at least $c$, i.e.\ $v$ is in $H_c$ if and only if:
\begin{align*}
    H(v) =&\; - \sum_{i=1}^n v_i \log(v_i) \geq  c.
\end{align*}
\end{definition}
We say that a constraint $H_c$ is {\it $\epsilon$-satisfied} by a  distribution $v$ if we have that $\min_{x \in H_c}d_{TV}(x, v) \leq \epsilon$, where $d_{TV}$ is the total variation distance between probability distributions.

Our algorithmic results can be extended to any convex constraint set which contains a small region around the uniform distribution, but we focus on entropy constraints as they are quite natural and have interesting connections to our setting which we consider in Section \ref{subsec:model-conditions}.

\subsection{Recommendation Menus for Adaptive Agents}
An instance of our problem consists of an item set $N = [n]$, a menu size $k$,  a preference model $M$ for the Agent, a constraint $H_c$, a horizon length of $T$ rounds, and a sequence of linear reward functions $\rho_1, \ldots, \rho_T$ for the Recommender. In each round $t \in \{1,\ldots,T\}$:
\begin{itemize}
    \item The Recommender chooses a menu $K_t \subset N$ with $\abs{K_t} = k$.  
    \item The Agent chooses item $i \in K_t$ with probability 
    \begin{align*}
        p_{K_t, v_t, i_t} =&\; \frac{s_{v_t, i_t}}{\sum_{j \in K_t} s_{v_t, j}}
    \end{align*}
    and updates its memory vector to the normalized histogram $$v_{t+1} = \frac{e_{i}}{t+1} + \frac{t \cdot v_t}{t+1},$$
    where $e_{i}$ is the $i$th standard unit vector.
    \item The Recommender observes receives reward $\rho_{t}(e_{i})$ for the chosen item.
\end{itemize}
The goal of the Recommender is to maximize their reward over $T$ rounds subject to $v_T$ satisfying $H_c$.
It might seem to the reader that the Recommender can `manipulate'
the Agent to achieve any preference score  vector over time; however, this is not true as many score vectors might not be achievable depending on the preference model. 

\subsection{Realizability Conditions for Item Distributions}

For any memory vector $v$, we define the feasible set of item choice distributions for Agent in the current round, each generated by a distribution over menus which the Recommender samples from.
\begin{definition}[Instantaneously-Realizable Distributions at $v$]
Let $p_{K,v} \in \Delta(n)$ be the item distribution selected by an Agent presented with menu $K$ at memory vector $v$, given by:
\begin{align*}
    p_{K,v,i} = \frac{s_{v,i}}{\sum_{j \in K} s_{v,j}}.
\end{align*}
The set of instantaneously-realizable distributions at $v$ is given by:
\begin{align*}
    \textup{\texttt{IRD}}(v, M) =&\; \convhull_{K \in {n \choose k}} p_{K, v}.
\end{align*}
\end{definition}
For any $x \in \textup{\texttt{IRD}}(v, M)$, any menu distribution $z \in \Delta({n \choose k})$ specifying a convex combination of menu score vectors $p_{K, v}$ which sum to $x$ will generate the item distribution $x$ upon sampling. 

One might hope to match the performance of the best menu distribution, or perhaps the best realizable item distribution from the uniform vector. Unfortunately, neither of these are possible.

\begin{theorem}\label{thm:regret-lowerbounds}
There is no algorithm which can obtain $o(T)$ regret against the best {item} distribution in the $\textup{\texttt{IRD}}$ set for the uniform vector, or against the best { menu} distribution in $\Delta\parens{{n \choose k}}$, even when the preference model is known exactly and is expressible by univariate linear functions.
\end{theorem}
We give a separate construction for each claim, with the full proof deferred to Appendix . 
The first is a case where the optimal distribution from the uniform vector cannot be played every round, as it draws the the memory vector into $\textup{\texttt{IRD}}$ sets where the reward opportunities are suboptimal. The second considers menu distributions where obtaining their late-round performance requires committing early to an irreversible course of action.
Instead, our benchmark will be the set of distributions which are realizable from {\it any} memory vector.

\begin{definition}[Everywhere Instantaneously-Realizable Distributions]
For a preference model $M$, the set of everywhere instantaneously-realizable distributions is given by:
\begin{align*}
    \textup{\texttt{EIRD}}(M) =&\; \bigcap_{v \in \Delta(n)} \textup{\texttt{IRD}}(v, M).
\end{align*}
This is the set of distributions $x \in \Delta(n)$ such that from any memory vector $v$, there is some menu distribution $z$ such that sampling menus from $d$ induces a choice distribution of $x$ for the agent.
\end{definition}
Note that the set $\texttt{EIRD}(M)$ is convex, as each $\texttt{IRD}(v, M)$ is convex by construction.

\subsection{Conditions for Preference Models}
\label{subsec:model-conditions}

The algorithm we present in Section \ref{sec:uniform} requires two key conditions for a class of preference models: each model in the class must be {\it dispersed}, and the class must be {\it locally learnable}.
This enforces that the Agent is always willing to select every item in the menu they see with some positive probability, and that the behavior at any memory vector can be estimated by observing behavior in a small region.

\begin{definition}[Dispersion]
A preference model $M$ is $\lambda$-dispersed if $s_{v, i} \geq \lambda$ for all $v \in \Delta(n)$ and for all $i$, i.e.\ items always have a score of at least $\lambda$ at any memory vector.
\end{definition}

The dispersion condition plays an important role in the analysis of our algorithm by enabling efficient exploration, but it additionally coincides with diversity constraints in appropriate regimes.

\begin{theorem}[High-Entropy Containment in \texttt{EIRD}]\label{thm:high-entropy-eird}
Consider the diversity constraint $H_{c}$ for $c = \log(n) - \gamma$, and let $\tau \geq \exp(-\gamma)$. Let $M$ be a $\lambda$-dispersed preference model with $\lambda \geq \frac{k^2 \exp(\gamma / \tau)}{n}$.
For any vector $v \in H_c$, there is a vector $v' \in \textup{\texttt{EIRD}}(M)$ such that $d_{TV}(v, v')$ 
is at most $O(\tau)$.
\end{theorem}

The key step here, proved in Appendix \ref{subsec:entropy-eird}, is that $\textup{\texttt{EIRD}}(M)$ contains the uniform distribution over any large subset of items, and taking mixtures of these can approximate any high-entropy distribution.

Next, for a class of models to be locally learnable, one must be able to accurately estimate a model's preference scores everywhere when only given access to samples in an arbitrarily small region.

\begin{definition}[Local Learnability]
Let $\M$ be a class of preference models, and let 
\begin{align*}
    \textup{\texttt{EIRD}}(\M) =&\; \bigcap_{M \in \M} \textup{\texttt{EIRD}}(M).
\end{align*}
Let $v^*$ be a point in $\textup{\texttt{EIRD}}(\M)$, and $V_{\alpha}$ be the set of points within 
distance $\alpha$ from $v^*$, for $\alpha$ such that $V_{\alpha} \subseteq \textup{\texttt{EIRD}}(\M)$. $\M$ is 
$h$-locally learnable
if there is some $v^*$ and an algorithm $\A$ which, for any $M \in \M$ and any $\alpha > 0$,
given query access to normalized score estimates $\hat{s}_v$ where $\norm{\hat{s}_v - M(v) / {M}_v^*}_{\infty} \leq \beta$ for any $v \in V_{\alpha}$ (where ${M}_v^* = \sum_i M(v)_i$) and for some $\beta$, can produce a hypothesis model $\hat{M}$ such that $\norm{\hat{M}(x)/\hat{M}^*_x - M(x)/M^*_x} \leq \epsilon$ for any $x \in \Delta(n)$ and $\epsilon = \Omega(\beta)$.
\end{definition}

The local learnability condition, while covering many natural examples shown in Section \ref{subsec:ll-models}, is indeed somewhat restrictive. In particular, it is not difficult to see that classes of piecewise functions, such as neural networks with ReLU activations, are not locally learnable. However, this appears to be essentially a necessary assumption for efficient learning, given the cumulative nature of memory in our setting. We show a runtime lower bound for any algorithm that hopes to learn an estimate $\hat{M}$ for the preference model $M$ via {\it queries}.
Even a Recommender who can force the Agent to pick a particular item each round, and exactly query the preference model for free at the current memory vector, may require exponentially many rounds to learn $\hat{M}$ if the points it must query are far apart.
\begin{theorem}[Query Learning Lower Bound]\label{thm:query-learning-lb}
Suppose the Recommender can force the Agent to select any item at each step $t$, and can query $M(v_t)$ at the current memory vector $v_t$. Let $\A_S$ be an algorithm which produces a hypothesis $\hat{M}$ by receiving queries $M(v)$ for each $v \in S$. For points $v$ and $v'$, let $d_{\max}(v, v') = \max_i v_i - v_i'$. Then, any sequence of item selections and queries by the Recommender requires at least
\begin{align*}
    T \geq&\; \min_{\sigma \in \pi(S)} \prod_{i=1}^{|S|- 1} (1 + d_{\max}(\sigma(i), \sigma(i+1)) )
\end{align*}
rounds to run $\A(S)$, where $\pi(S)$ is the set of permutations over $S$ and $\sigma(i)$ is the $i$th item in $\sigma$.
\end{theorem}

We prove this in Appendix \ref{subsec:query-lb}. Notably, this implies that if $S$ contains $m$ points which, for any pair $(v, v')$ have both $d_{\max}(v, v') \geq \gamma$ and $d_{\max}(v', v)\geq \gamma$, at least $O\parens{(1 + \gamma)^m}$ rounds are required.

\subsection{Locally Learnable Preference Models}
\label{subsec:ll-models}

There are several interesting examples of model classes which are indeed locally learnable, which we prove in Appendix \ref{sec:proofs-ll}. In general, our approach is to query a grid of points inside the radius $\alpha$ ball around the uniform vector, estimate each function's parameters and show that the propagation of over the entire domain is bounded. Note that the normalizing constants for each query we observe may differ; for univariate functions, we can handle this by only moving a subset of values at a time, allowing for renormalization. For multivariate polynomials, we consider two distinct classes and give a separate learning algorithm for each; we can estimate ratios of scores directly for multilinear functions, and if scores are already normalized we can avoid rational functions altogether. 
Each local learning result we prove involves an algorithm which makes queries near the uniform vector. We later show in Lemma \ref{lemma:unif-eird} that taking $\lambda \geq k^2/n$ suffices to ensure that these queries can indeed be implemented via an appropriate sequence of menu distributions for any $M$ in such a class.

\subsubsection{Bounded-Degree Univariate  Polynomials}

Let $\M_{BUP}$ be the class of  {\it bounded-degree univariate polynomial} preference models where:

\begin{itemize}
    \item For each $i$, $M(v)_i = f_i(v_i)$, where $f_i$ is a degree-$d$ univariate polynomial which takes values in $[\lambda, 1]$ over the range $[0,1]$ for some constant $\lambda > 0$. 
\end{itemize}

Univariateness captures cases where relative preferences for an item depend only on the weight of that item in the agent's memory, i.e. there are no substitute or complement effects between items.

\begin{lemma}
$\M_{BUP}$ is 
$O(d)$-locally learnable
by an algorithm $\A_{BUP}$ with  $\beta \leq O(\epsilon \lambda^2 \cdot (\frac{\alpha}{nd})^d)$. 
\end{lemma}

\subsubsection{Bounded-Degree Multivariate Polynomials}

Let $\M_{BMLP}$ be the class of  {\it bounded-degree multilinear polynomial} preference models where:
\begin{itemize}
    \item For each $i$, $M(v)_i = f_i(v)$, where $f_i$ is a degree-$d$ multilinear (i.e.\ linear in each item) polynomial which takes values in $[\lambda, 1]$ over $\Delta(n)$  for some constant $\lambda > 0$, 
\end{itemize}
and let $\M_{BNMP}$ be the class of {\it bounded-degree normalized multivariate polynomial} preference models where:
\begin{itemize}
    \item For each $i$, $M(v)_i = f_i(v)$, where $f_i$ is a degree-$d$ polynomial which takes values in $[\lambda, 1]$ over $\Delta(n)$  for some constant $\lambda > 0$, where $\sum_i f_i(v) = C$ for some constant $C$. 
\end{itemize}

Together, these express a large variety of adaptivity patterns for preferences which depend on frequencies of many items items simultaneously. In particular, these can capture relatively intricate ``rabbit hole'' effects, in which some subsets of items are mutually self-reinforcing, and where their selection can discourage future selection of other subsets.

\begin{lemma}
$\M_{BMLP}$ and $\M_{BNMP}$ are both
${O}({n^d})$-locally learnable, for $\beta \leq {O}(\frac{\epsilon^2 }{ \textup{poly}(n (d/ \alpha)^{d})) })$, and $\beta \leq \frac{\epsilon}{\alpha^{d} F(n, d)} $, respectively, where $F(n,d)$ is independent of other parameters.
\end{lemma}

\subsubsection{Univariate Functions with Sparse Fourier Representations}
We can also allow for classes of functions where the minimum allowable $\alpha$ depends on some parameter. Functions with sparse Fourier representations are such an example, and naturally capture settings where preferences are somewhat cyclical, such as when an Agent goes through ``phases'' of preferring some type of content for a limited window.
We say that a function $f : \R \rightarrow \R$ is \emph{$\ell$-sparse}
if $f(x) = \sum_{i=1}^\ell \xi_i e^{2\pi \mathbf{i} \eta_i x}$ where $\eta_i \in [-F,F]$ denotes the $i$-th frequency 
and $\xi_i$ denotes the corresponding magnitude.
We say that an $\ell$-sparse function $f$ is $\hat{\alpha}$-separable when $\min_{i \neq j} |\eta_i - \eta_j| > \hat{\alpha}$.
Let $\M_{SFR}(\hat{\alpha})$ be the class of univariate {\it sparse Fourier representation} preference models where:
\begin{itemize}
    \item For each $i$ , $M(v)_i = f_i(v_i)$, where $f_i$ is a univariate  $\ell$-sparse and $\hat{\alpha}$-separable function which, over $[0,1]$,  
    is $L$-Lipschitz and takes values in $[\lambda, 1]$ for some constant $\lambda > 0$. 
\end{itemize}

\begin{lemma}
$\M_{SFR}(\hat{\alpha})$ is 
$\widetilde{O}(n \ell)$-locally learnable by an algorithm $\A_{SFR}$ with $\beta \leq O(\frac{\epsilon \lambda \alpha}{\sqrt{n} \ell} )$ and any $\alpha \geq \Tilde{\Omega}(1/\hat{\alpha})$.
\end{lemma}

%% file: contracting.tex
\section{Bandit Linear Optimization with Contracting Sets}
\label{sec:contracting}

The inner problem for the Recommender can be viewed as a bandit linear optimization problem over $H_c \cap \texttt{EIRD}(M)$. However, representing $\texttt{EIRD}(M)$ is challenging even if we know $M$ exactly, as it involves an intersection over infinitely many sets (generated by a possibly non-convex function), and a 
net approximation 
would involve exponential dependence on $n$.
Instead, our approach will be to operate over the larger set $H_c \cap \parens{ \bigcap_{t} \texttt{IRD}(v_t, M)}$ for the memory vectors $v_t$ we have seen thus far, where representing each $\texttt{IRD}$ has exponential dependence only on $k$
(from enumerating all menus). 

The tradeoff is that we can no longer directly use off-the-shelf bandit linear optimization algorithms for a known and fixed decision set such as \textsc{FKM} \cite{DBLP:journals/corr/cs-LG-0408007} or \textsc{SCRiBLe} \cite{scrible} as a subroutine, as our decision set is contracting each round. We introduce an algorithm for bandit linear optimization, a modification of the \textsc{FKM} algorithm we call Robust Contracting \textsc{FKM} (\textsc{RC-FKM}), which handles this issue by projecting to our estimate of the contracted decision set at each step. Additionally, \textsc{RC-FKM} can handle the imprecision resulting from our model estimation step, which can be represented by small adversarial perturbations to the action vector in each round; we modify the sampling rule to ensure that our target action remains in the true decision set even when perturbations are present. We prove the regret bound for \textsc{RC-FKM} in Appendix \ref{sec:rcfkm-proofs}.

\begin{algorithm}
\caption{(Robust Contracting FKM).}\label{alg:bandit-constricting}
\begin{algorithmic}
\State Input: sequence of contracting convex decision sets $\K_1, \ldots \K_T$ containing $\mathbf{0}$, perturbation vectors $\xi_1,\ldots , \xi_T$ where $\norm{\xi_t} \leq \epsilon$, parameters $\delta$, $\eta$.
\State Set $x_1 = \mathbf{0}$
\For{$t = 1$ to $T$}
    \State Draw $u_t \in \mathbb{S}_1$ uniformly at random, set $y_t = x_t + \delta u_t + \xi_t$
    \State Play $y_t$, observe and incur loss $\phi_t \in [0,1]$, where  $\E[\phi_t] = f_t(y_t)$ 
    \State Let $g_t = \frac{n}{\delta} \phi_t u_t$
    \State Let $\K_{t+1,\delta, \epsilon} =\{x \vert \frac{r}{r-\delta - \epsilon} x_t \in \K_{t+1} \}$
    \State Update $x_{t+1} = \Pi_{\K_{t+1,\delta,\epsilon}}[x_t - \eta g_t]$ 
\EndFor
\end{algorithmic}
\end{algorithm}

\begin{theorem}[Regret Bound for Algorithm \ref{alg:bandit-constricting}]
For a sequence of $G$-Lipschitz linear losses $f_1,\ldots,f_T$ and a contracting sequence of domains $\K_1,\ldots,\K_T$ (with $\K_j \subseteq \K_i$ for $j > i$, each with diameter at most $D$, and where a ball of radius $r > \delta+\epsilon$ around $\mathbf{0}$ is contained in $\K_T$), and adversarially chosen unobserved vectors $\xi_1,\ldots , \xi_T$ with $\norm{\xi_t} \leq \epsilon$ which perturb the chosen action at each step,
with parameters $\eta = \frac{D}{nT^{3/4}}$ and $\delta = \frac{1}{T^{1/4}}$,
Algorithm \ref{alg:bandit-constricting} obtains the expected regret bound
\begin{align*}
    \sum_{t=1}^{T} \E[\phi_t] - \min_{x \in \K_T} \sum_{t=1}^T f_t(x) \leq&\; n GDT^{3/4} + \frac{ G D T^{3/4}}{r} +  \frac{2 \epsilon GDT}{r}.
\end{align*}
\end{theorem}

%% file: uniform.tex
\section{Recommendations for Adaptive Agents }
\label{sec:uniform}

Our main algorithm begins with an explicit learning phase, after which we conduct regret minimization, and at a high level works as follows:
\begin{itemize}
    \item First, we learn an estimate of the preference model $\hat{M}$ by implementing local learning with a set of points close to the uniform memory vector, which suffices to ensure high accuracy of our representation with respect to $M$. If the number of local learning queries is independent of error terms and $\beta = \Theta(\epsilon)$, we can complete this stage in $t_0 = \tilde{O}(1/\epsilon^3) = \tilde{O}(T^{3/4})$ steps.
    \item For the remaining $T - t_0$ steps, we implement \textsc{RC-FKM} by using the learned model $\hat{M}$ at each step to solve for a menu distribution which generates the desired item distribution from the current memory vector, then contracting the decision set based on the memory update.
\end{itemize}

\begin{theorem}[Regret Bound for Algorithm \ref{alg:unif-no-regret}]\label{thm:unif-regret}
Algorithm \ref{alg:unif-no-regret} obtains regret bounded by
\begin{align*}
    \textup{Regret}_{C \cap \textup{\texttt{EIRD}}(M)}(T) \leq&\; \Tilde{O} \parens{ t_0 + n GT^{3/4} + \frac{(\delta + \epsilon) G T}{r} + \epsilon GT} = \tilde{O}(T^{3/4}) 
\end{align*}
where $t_0$ is the time required for local learning,
$r = O(k^2 / n)$, and $\epsilon, \delta = O(r \cdot T^{-1/4})$, and results in an empirical distribution such that $H_c$ is $O(\epsilon)$-satisfied with probability at least $1 - O(T^{-1/4})$.
\end{theorem}

\begin{algorithm}
\caption{A no-regret recommendation algorithm for adaptive agents.}\label{alg:unif-no-regret}
\begin{algorithmic}
\State Input: Item set $[n]$, menu size $k$, Agent with $\lambda$-dispersed memory model $M$ for $\lambda \geq \frac{k^2}{n}$, where $M$ belongs to an $S$-locally learnable class $\M$, diversity constraint $H_c$, horizon $T$, $G$-Lipschitz linear losses $\rho_i, \ldots, \rho_T$.
\State Let $t_{\text{pad}} = \Tilde{\Theta}(1/\epsilon^3)$
\State Let $t_{\text{move}} = \Tilde{\Theta}(1/\epsilon^3)$
\State Let $t_{\text{query}} = \Tilde{\Theta}(1/\epsilon^2)$
\State Let $\alpha = \Theta( \frac{k}{n^2 S} )$
\State Get set of $S$ points in the $\alpha$-ball around uniform vector $x_U$ to query from $\A_{\mathcal{M}}$
\State Let $t_0 = t_{\text{pad}} + S(2 \cdot t_{ \text{move}} + t_{\text{query}})$
\State Run $\texttt{UniformPad}$ for $t_{\text{pad}}$ rounds
\For{$x_i$ in $S$}
\State Run $\texttt{MoveTo}(x_i)$ for $t_{\text{move}}$ rounds
\State Run $\texttt{Query}(x_i)$ for $t_{\text{query}}$ rounds, observe result $\hat{q}(x_i)$
\State Run $\texttt{MoveTo}(x_U)$ for $t_{\text{move}}$ rounds
\EndFor
\State Estimate model $\hat{M}$ using $\A_{\M}$ for $\beta = \Theta(\epsilon)$
\State Let $v_{t_0}$ be the empirical item distribution of the first items $t_0$ items
\State Let $\K_{t_0} = H_c$ (in $n-1$ dimensions, with $x_{t,n} = 1 - \sum_{i=1}^{n-1} x_{t,i}$, and s.t.\ $x_U$ translates to $\mathbf{0}$)
\State Initialize $\textsc{RC-FKM}$ to run for $T^* = T - t_0 - 1$ rounds with $r = O(k^2 / n)$, $\delta, \epsilon = \frac{r}{{T^*}^{1/4}}$
\For{$t = t_0 + 1$ to $T$}
\State Let $x_t$ be the point chosen by \textsc{RC-FKM} 
\State Use $\texttt{PlayDist}(x_t)$ to compute menu distribution $z_t$
\State Sample $K_t \sim z_t$, show $K_t$ to Agent
\State Observe Agent's chosen item $i_t$ and reward $\rho_t(e_{i_t})$
\State Update \textsc{RC-FKM} with $\rho_t(e_{i_t})$
\State Let $v_{t} = \frac{t-1}{t}v_{t-1} + \frac{1}{t} \cdot e_{i_t}$
\State Update the decision set to $\K_{t+1} = \K_t \cap ~ \texttt{IRD}(v_t, \hat{M})$
\EndFor
\end{algorithmic}
\end{algorithm}

\subsection{Structure of $\texttt{EIRD}(M)$}
The key tool which enables us to implement local learning is a construction for generating any point near the uniform via an adaptive sequence of menu distributions, provided $\lambda$ is sufficiently large.

\begin{lemma}\label{lemma:unif-eird}
For any $\lambda$-dispersed $M$ where $\lambda \geq \frac{k^2}{n}$, $\textup{\texttt{EIRD}}(M)$ contains all points $x \in \Delta(N)$ satisfying 
\begin{align*}
    \norm{x - x_U}_{\infty} \leq&\; \frac{k-1}{n(n-1)},
\end{align*}
where $x_U$ is the uniform $\frac{1}{n}$ vector.
\end{lemma}

We give an algorithmic variant of this lemma which is used directly by Algorithm \ref{alg:unif-no-regret}, as well as a variant for uniform distributions over smaller subsets as $\lambda$ grows, which we use to prove Theorem \ref{thm:high-entropy-eird}. 

\subsection{Subroutines}
 
Our algorithm makes use of a number of subroutines for navigating the memory space, model learning, and implementing \textsc{RC-FKM}. We state their key ideas here, with full details deferred to Appendix \ref{sec:main-alg-proof}.
 
\paragraph{\texttt{UniformPad}:}
\begin{itemize}
    \item In each round, include the $k$ items with smallest counts, breaking ties randomly.
\end{itemize}
 
\paragraph{\texttt{MoveTo}($x$):}
\begin{itemize}
    \item Apply the same approach from \texttt{UniformPad} to the difference between the current histogram and $x$.
\end{itemize}

\paragraph{\texttt{Query}($x$):}
\begin{itemize}
    \item Play a sequence of $O(n/k)$ partially overlapping menus which cover all items, holding each constant long enough for concentration, and compute relative probabilities of each item.
\end{itemize}

\paragraph{\texttt{PlayDist}($x$):}
\begin{itemize}
    \item Given an item distribution $x$, we solve a linear program to compute a menu distribution $z_x$ using $\hat{M}(v)$ which induces $x$ when a menu is sampled and the Agent selects an item.
\end{itemize}

The intuition behind our learning stage is that each call to \texttt{Query}($x$) can be accurately estimated by bounding the ``drift'' in the memory vector while sampling occurs, as the number of samples per query is small compared to the history thus far. Each call to \texttt{MoveTo}($x$) for a point within the $\alpha$-ball can be implemented by generating an empirical distribution corresponding to a point in $\texttt{EIRD}(M)$ for sufficiently many rounds.
 
The resulting model estimate $\hat{M}$ yields score estimates which are accurate for any memory vector. To run \textsc{RC-FKM}, we translate to an $n-1$ dimensional simplex representation, and construct a menu distribution to implement any action $x_t$ via a linear program (\texttt{PlayDist}($x$)). The robustness guarantee for \textsc{RC-FKM} ensures that the loss resulting from imprecision in $\hat{M}$ is bounded, and further ensures that the resulting expected distribution remains inside $H_c$ (and that $H_c$ is approximately satisfied with high probability by the empirical distribution).
We contract our decision set in each step with the current space $\texttt{IRD}(v_t, \hat{M})$, which will always contain $\texttt{EIRD}(\hat{M})$, the best point in which is competitive with the best point in $\texttt{EIRD}({M})$.

%% file: future.tex
\section{Conclusion and Future Work}
\label{sec:conclusion}

Our work formalizes a bandit setting for investigating online recommendation problems where agents' preferences can adapt over time and provides a number of key initial results which highlight the importance of diversity in recommendations,
including lower bounds for more ``ambitious'' regret benchmarks, and a no-regret algorithm for the \texttt{EIRD} set benchmark, which can coincide with the high-entropy set under appropriate conditions.  
Our results showcase a tradeoff between the space of strategies one considers and the ability to minimize regret. Crucially, our lower bound constructions illustrate that we cannot hope to optimize over the set of recommendation patterns which may send agents down ``rabbit holes'' that drastically alter their preferences, whereas it is indeed feasible to optimize over the space of sufficiently diversified recommendations.

There are several interesting directions which remain open for future investigation,
including additional characterizations of the \texttt{EIRD} set, 
discovering more examples or applications for local learnability, 
identifying the optimal rate of regret or dependence on other parameters, settings involving multiple agents with correlated preferences, and consideration of alternate models of agent behavior which circumvent the difficulties posed by uniform memory.

%% file: appendix.tex
\section{Further Related Work}
\label{sec:app-related}
\subsection{Empirical Investigation Of Recommendation Feedback Loops}

A substantial body of evidence has emerged in recent years indicating that recommendation systems can create feedback loops which drive negative social consequences. \cite{o2015down} observed that users accessing videos with extreme political views are likely to get caught in an ``ideological bubble'' in just a few clicks, and \cite{10.1145/3308560.3317596} explore the role of recommendation algorithms in creating distrust and amplifying political polarization on social media platforms.
By investigating a real-world e-commerce dataset, \cite{DBLP:journals/corr/abs-2007-02474} study the way in which recommendation systems drive agents' self-reinforcing preferences and lead them into ``echo chambers'' where they are separated from observing a diversity of content. \cite{10.1145/3442381.3450099} conduct a meta-analysis over many datasets which focuses specifically on the ``rabbit hole'' problem by means of exploring ``taste distortion'' of agents who observe recommendations which are more extreme than their current preferences. Such results motivate investigating these dynamics from game-theoretic and learning-theoretic foundations.

\subsection{Modeling Feedback Loops in Recommendation Systems}

A number of recent works from the recommendation systems literature have explored the role of collaborative filtering algorithms for various models of agent behavior, aiming to understand how feedback loops in recommendation patterns emerge, the harms they cause, and how they can be corrected \cite{DBLP:journals/corr/abs-2201-07203, 10.5555/3157382.3157461}.
A common theme is homogenization of recommendations across a {\it population} of users, which can lead to exacerbation of biased utility distributions for minority groups \cite{DBLP:journals/corr/abs-2007-13019}, long-run utility degradation \cite{DBLP:journals/corr/abs-1710-11214}, and a lack of traffic to smaller content providers which results in them being driven to exit the platform \cite{DBLP:journals/corr/abs-2008-00104}.
Our work indirectly addresses this phenomenon by encouraging diverse recommendations, but our primary focus is from the perspective of a single agent, who may be led down a ``rabbit hole'' by an algorithm which optimizes for their immediate engagement.

\subsection{Dueling Bandits}

The ``dueling bandits'' problem, initially proposed as a model for similar recommendation systems challenges \cite{10.1145/1553374.1553527, YUE20121538}, and which has been generalized for sets larger than two \cite{NEURIPS2020_d5fcc35c, https://doi.org/10.48550/arxiv.2101.01572}, considers a similar setting in which bandit optimization is conducted with respect to the preference model of an agent, occasionally represented via an explicit parametric form. Here, one presents a set of choices to an agent, then receives only {\it ordinal} feedback about the relative rewards of the choices, and must optimize recommendations with regret measured against the best individual choice. In contrast to our setting, these works consider preferences which are fully determined {\it a priori}, and do not change as a function of item history or exhibit preference feedback loops.

\subsection{Online Stackelberg Problems}

A number of works in recent years explore online problems where an agent responds to the decision-maker's actions, influencing their reward. The performative prediction setting, introduced in \cite{DBLP:journals/corr/abs-2002-06673}, captures settings in which a deployed classifier results in changes to the distribution itself, in turn affecting performance.
This work has been extended to handle stochastic feedback \cite{NEURIPS2020_33e75ff0} and notably, to a no-regret variant \cite{DBLP:journals/corr/abs-2202-00628} which involves learning mapping between classifiers and distribution shifts, which bears some conceptual similarities to our procedure for locally learning an agent's preference model. The ``revealed preferences'' literature involves a similar requirement of learning a mapping between actions and agent choices \cite{DBLP:journals/corr/RothUW15, DBLP:journals/corr/abs-1710-07887}.
Some features of our setting resemble elements of other well-studied online problems, including the restricted exploration ability for limited switching problems (e.g.\ \cite{pmlr-v75-altschuler18a}), and the contracting target set for chasing nested convex bodies (e.g.\ \cite{doi:10.1137/1.9781611975994.91}).

\subsection{Strategizing Against Adaptive Agents}

Some recent work has begun to explore the problem of designing optimal strategies in a repeated game against agents who {\it adapt} their strategies over time using a no-regret algorithm. In auction problems, \cite{DBLP:journals/corr/abs-1711-09176} study the extent to which an auction designer can extract value from bidders who use different kinds of no-regret algorithms.
More generally, \cite{DBLP:journals/corr/abs-1909-13861} connect this line of investigation to Stackelberg equilibria for normal-form games.
In strategic classification problems, \cite{Zrnic2021WhoLA} study the behavior when using a learning rate which is either much faster or much slower than that of the agents which one aims to classify, and draw connections to equilibrium concepts as well.  Our work extends this notion of strategizing against adaptive agents to recommendation settings, with novel formulations of adaptivity and regret to suit the problem's constraints.

\section{Omitted Proofs for Sections \ref{sec:prelims}}

\subsection{Proof of Linear Regret Lower Bounds (Theorem \ref{thm:regret-lowerbounds})}
\label{subsec:lower-bound-proofs}

We give a separate lower bound construction for the uniform $\texttt{IRD}$ item distribution benchmark and the menu distribution benchmark, yielding the theorem.

\begin{lemma}
There is no algorithm which can obtain $o(T)$ regret against the best {item} distribution in the $\textup{\texttt{IRD}}$ set for the uniform vector, even when the preference model is known exactly and is expressible by univariate linear functions.
\end{lemma}

\begin{proof}
First we give an example for which obtaining $o(T)$ regret against $\textup{\texttt{IRD}}(v, M)$ for the uniform vector $v_U$ is impossible.
Consider the memory model $M$ where:
\begin{itemize}
    \item $M(v)_1 = \lambda + 0.5 + \frac{n}{n-1}\cdot (v_1 - \frac{1}{n})\cdot(0.5 - \lambda)$;
    \item $M(v)_2 = \lambda +  0.5 (1 - v_1 + \frac{1}{n})$; 
    \item $M(v)_i = 0.5 + \lambda$ for $i > 2$.
\end{itemize}
Observe that at the uniform distribution where $v_1 = \frac{1}{n}$, all items have a score of $0.5 + \lambda$.
If $v_1 = 1$, we have that:
\begin{itemize}
    \item $M(v)_1 = 1$, and 
    \item $M(v)_2 = \lambda + \frac{0.5 }{n}$.
\end{itemize}
If $v_1 = 0$, we have that:
\begin{itemize}
    \item $M(v)_1 = \lambda + 0.5 - \frac{0.5 - \lambda}{n-1}$~, and
    \item $M(v)_2 = \lambda + 0.5 \cdot(1 + \frac{1}{n})$ 
\end{itemize}
As scores linearly interpolate between these endpoints for any $v_1$, $M$ is $\lambda$-dispersed, and scores lie in $[\lambda, 1]$.
Let $k=2$. Consider reward functions which give reward $\alpha > 0$ for item 1 in each round up to $t^* = T/2$, giving reward 0 to each other item; after $t^*$, a reward of $\beta > 0$ is given for item $2$ while the rest receive a reward $0$. The distribution which assigns probability $1/2$ each to item 1 and 2, with all other items having probability 0, is contained in $\textup{\texttt{IRD}}(v_U, M)$, as one can simply play the menu with both items. This distribution yields a total expected reward of
\begin{align*}
    R_v =&\; \frac{\alpha t^*}{2} + \frac{\beta t^*}{2}
\end{align*}
over $T$ steps. Consider the performance of any algorithm $\A$ which results in item 1 being selected with an empirical probability $p$ 
over the first $t^*$ rounds.
At $t=t^*$, we have $v_{t^*,1} = p$; its total reward over the first $t^*$ rounds is $\alpha p t^*$.
For sufficiently large $n$ and small $\lambda$, the score for item 2 is approximated by
$M(v)_2 = 0.5(1 -p)$ up to any desired accuracy. In future rounds $t \geq t^*$, the value $v_{t,s1}$ is at least
$\frac{pt^*}{t}$, 
and so the score for item 2 is at most
\begin{align*}
    M(v)_2 =&\; 0.5(1 - \frac{pt^*}{t}).
\end{align*}
Each other item has a score of at least $0.5$, yielding an upper bound on the probability that item 2 can be selected even if it is always in the menu, as well as a maximum expected per-round reward of
\begin{align*}
    R_t =&\; \beta \cdot\parens{\frac{0.5(1 - \frac{pt^*}{t})}{1 - 0.5\frac{pt^*}{t}  }}.
\end{align*}

At time $T = 2 t^*$, the instantaneous reward is at most 
\begin{align*}
    R_T =&\; \beta \cdot\parens{\frac{2 - p}{4 - p  }},
\end{align*}
which is also a per-round upper bound for each $t \geq t^*$. This bounds the total reward for $\A$ by
\begin{align*}
    R_{\A} =&\; \alpha p t^* + \beta t^*  \cdot \frac{2 - p}{4 - p}.
\end{align*}
We can now show that for any $p$, there exists a $\beta$ such that $R_v - R_{\A} = \Theta(T)$. For any $p \leq \frac{1}{3}$, we have 
\begin{align*}
    R_{\A} \leq &\; \frac{ \alpha t^* }{3} + \frac{\beta t^* }{2},
\end{align*}
and for any $p > \frac{1}{3}$ we have:
\begin{align*}
    R_{\A} \leq &\; { \alpha t^* } + \frac{5\beta t^* }{11}.
\end{align*}
In the first case, we immediately have $R_v - R_{\A} \geq T \alpha / 6$ for any $\beta$. In the second case, let $\beta \geq 22 \alpha$. We then have: 
\begin{align*}
    R_v - R_{\A} \geq&\; \frac{\beta t^* }{22} - \frac{ \alpha t^* }{2}  \\ 
    \geq&\; T \alpha / 4. 
\end{align*}
The value of $\beta$ can be determined adversarially, and so there is no algorithm $\A$ which can obtain $o(T)$ regret against $\textup{\texttt{IRD}}(v, M)$.

\end{proof}

Next we show a similar impossibility result for regret minimization with respect to the set of all menu distributions.

\begin{lemma}
There is no algorithm which can obtain $o(T)$ regret against the best  menu distribution in $\Delta\parens{{n \choose k}}$, even when the preference model is known exactly and is expressible by univariate linear functions.
\end{lemma}

\begin{proof}
Let $M$ be the $\lambda$-dispersed memory model where 
the functions for items $(a,b,c)$, and every other item $i$,
are given by:
\begin{itemize}
    \item $M(v)_a = \lambda + (1 - \epsilon) (1 - v_b)$;
    \item $M(v)_b = \lambda + (1 - \epsilon) v_b$;
    \item $M(v)_c = \lambda + (1 - \epsilon) v_c$;
    \item $M(v)_i = \lambda + (1 - \epsilon) (1 - v_b)$~ for $i \notin \{a, b, c\}$; 
\end{itemize}
for some $\lambda > 0$ and $\epsilon > \lambda$. Let $k=2$. 
Consider a sequence of rewards $\{f_t\}$ which yields reward $\alpha$ to items $(a,b)$ for each round $t \leq t^*$ and 0 to the rest, then in each step after $t^*$, yields 
a reward of $\beta$ for item $c$, a reward of $0$ for item $b$, and  reward of $-\beta$ for every other item.
Note the total expected reward for the following distributions:
\begin{align*}
    R_{(a,b)}(T) =&\; \alpha t^*  - \beta(T - t^*)/2 ; \\
    R_{(b,c)}(T) =&\; \alpha  t^*/2 + \beta(T - t^*)/2 ;
\end{align*}
The bound for $R_{(a,b)}(t^*)$ follows from symmetricity of the resulting stationary distribution, given by the unique solution $v_a = 0.5$ to the recurrence: 
$$v_b = \frac{\lambda + (1 - \epsilon)v_b}{2\lambda + (1 - \epsilon)}$$
which is approached in expectation for large $T$ regardless of initial conditions for any constant $\lambda$. Symmetricity also results in balanced expectations for each item in $R_{(b,c)}$.

Consider the distribution $p_{t^*}$ played by an algorithm $\A$ over the first $t^*$ rounds, where $t^*$ is large enough to ensure concentration. If $p_{t^*,a} + p_{t^*, b} \leq 1 - \delta$ for some constant $\delta$, then for $\beta = 0$ the algorithm has regret $\delta \alpha t^* = \Theta(T)$ for any $t^* = \Theta(T)$. Further, if regret is not bounded, the menu $(a, b)$ must be played in nearly every round, as other item placed in the menu has positive selection probability.
As such, the empirical probability of $b$ must be close to $1/2$. 

After $t^*$, the algorithm cannot obtain a per-round utility which matches that of $(b,c)$ up to $\delta$ until a round $t$ where either:
\begin{align*}
    \frac{\lambda + (1 - \epsilon) p_{t,c} }{2\lambda + (1 - \epsilon)(p_{t,b} + p_{t,c})} \geq&\; 1/2 - \delta
\end{align*}
or
\begin{align*}
    \frac{\lambda + (1 - \epsilon) p_{t,c} }{2\lambda + (1 - \epsilon)(1 - p_{t,b} + p_{t,c})} \geq&\; 1/2 - \delta,
\end{align*}
which requires the total number of rounds in which $c$ is chosen to approach $t^* / 2 - C \cdot \delta t^*$, where $C$ is a constant depending on $\epsilon$ and $\lambda$. Let $T = 3t^*/2$, and so this cannot happen for small enough constant $\delta$, resulting in a regret of $\delta \beta T / 3 - \alpha T/3 $ with respect to $(b,c)$, which is $\Theta(T)$ when $\delta \beta > \alpha $. 

\end{proof}

\subsection{Proof of High-Entropy Containment of \texttt{EIRD} (Theorem \ref{thm:high-entropy-eird})}
\label{subsec:entropy-eird}

\begin{proof}

By Lemma \ref{lemma:unif-subset-eird}, for a $\lambda$-dispersed preference model $M$ with $\lambda \geq \frac{Ck^2}{n}$, any uniform distribution over $n/C$ items lies inside $\textup{\texttt{EIRD}}(M)$.
We make use of a lemma from \cite{pmlr-v65-agarwal17c}, which we restate here.
\begin{lemma}[Lemma 8 in \cite{pmlr-v65-agarwal17c}]
For a random variable $A$ over $[n]$ with $H(A) \geq \log n - \gamma$, there is a set of $\ell + 1 = O(\gamma / \tau^3))$ distributions $\psi_i$ for $i \in \{0,\ldots,n\}$ over a partition of the support of $A$ which can be mixed together to generate $A$, where $\psi_0$ has weight $O(\tau)$, and where for each $i\geq 1$:
\begin{enumerate}
    \item $\log \abs{ \textup{supp}(\psi_i)} \geq \log n  - \gamma / \tau$.
    \item $\psi_i$ is within total variation distance $O(\tau)$ from the uniform distribution on its support.
\end{enumerate}
\end{lemma}
Using this, we can explicitly lower bound the support of each $\psi_i$:
\begin{align*}
    \log \abs{ \text{supp}(\psi_i) } \geq&\; \log(n) - \gamma/\tau \\ 
    =&\; \log(n) - \log(\exp(\gamma / \tau)) \\
    =&\; \log\parens{\frac{n}{\exp(\gamma/ \tau)}}.
\end{align*}
As such:
\begin{align*}
        \abs{\text{supp}(\psi_i) } \geq&\; \frac{n}{\exp(\gamma / \tau)}.
\end{align*}
Each uniform distribution over $ \text{supp}(\psi_i)$ lies inside $\textup{\texttt{EIRD}}(M)$ for $\lambda \geq \frac{Ck^2}{n}$, provided that $C \geq \exp(\gamma / \tau)$.
The $O(\tau)$ bound on total variation distance is preserved under mixture, as well as when redistributing the mass of $\psi_0$ arbitrarily amongst the uniform distributions.

\end{proof}

\subsection{Proof of Query Learning Runtime Lower Bound (Theorem \ref{thm:query-learning-lb})}
\label{subsec:query-lb}
\begin{proof}

For any permutation $\sigma$, we can lower bound the steps required to move between any two vectors adjacent in the ordering in terms of $d_{\max}$ and the number of rounds elapsed thus far. 

\begin{lemma}
Consider two vectors $v$ and $v'$, where $v$ is the current empirical item distribution after $t$ steps. Reaching an empirical distribution of $v'$ requires at least $t \cdot d_{\max}(v, v')$ additional steps.
\end{lemma}

\begin{proof}

Let $x$ be the histogram representation of $v$ with total mass $t$, and let $j^* = \text{arg max}_j v_j - v_j'$, where $v_j - v_j' = d_{\max}(v, v')$. Let $x' = t' \cdot v'$ be the histogram representation of $v'$ with total mass $t'$, such that $x_{j^*} = x_{j^*}'$. Note that $t'$ is the smallest total mass (or total number of rounds) where a histogram can normalize to $v'$, as any subsequent histogram must have $x_{j^*}' \geq x_{j^*}$. As such, we must have that $t'\cdot v_{j^*}' \geq t\cdot v_{j^*}$, implying that:
\begin{align*}
    \frac{t'}{t} \geq &\; \frac{v_{j^*}}{v_{j^*}'} \\
    =&\; \frac{v_{j^*}' + d_{\max}(v, v')}{v_{j^*}'} \\
    \leq&\; 1 + d_{\max}(v, v').
\end{align*}

\end{proof}

At least one round is required to reach the first vector in a permutation, and we can use the above lemma to lower-bound the rounds between any adjacent vectors in the ordering. Taking the minimum over all permutations gives us the result.
\end{proof}

\section{Proofs of Local Learnability for Section \ref{subsec:ll-models}}
\label{sec:proofs-ll}
Each proof gives a learning algorithm which operates in a ball around the uniform vector, which is contained in $\texttt{EIRD}(M)$ whenever $\lambda \geq \frac{k^2}{n}$ by Lemma \ref{lemma:unif-eird}.

\subsection{Proof of Univariate Polynomial Local Learnability}

\begin{proof}
Query the uniform vector $v_U$ where each $v_i=\frac{1}{n}$. 
Let $Z= \frac{\sqrt{n d / 6}}{\alpha} $.
Consider three sets each of $d/2$ memory vectors where the items with indices satisfying $i\mod 3 = z$ each have memory values $\frac{1}{n}+\frac{j}{Z}$, items satisfying $i \mod 3 = z+1$ have values $\frac{1}{n}-\frac{j}{Z}$, and the remainder have $\frac{1}{n}$ (for $z \in \{0,1,2\}$, and for $1 \leq j \leq  \frac{d}{2}$). 
All such vectors lie in $V_{\alpha}$, as $2n / 3 \cdot (d / (2Z))^2 \leq \alpha^2$.
Query each of the $3d/2$ vectors.
For each query,
let $R_v$ be the sum of all scores of the items held at $\frac{1}{n}$, divided by the sum of those same items' scores in the uniform query. Divide all scores by $R_v$.
Let $R_v^*$ be the be the corresponding ratio of these sums of scores under $\{ f_i \}$; each sum is within $[\frac{\lambda}{3}, 1]$ at each vector, and the sums of observed scores have additive error at most $n\beta/3$. As such, $R_v$ has additive error at most $\frac{2n\beta}{\lambda}$ from $R_v^*$.
This gives us estimates for $d+1$ points of $\hat{f}_i(x_j) = \hat{y}_j$  for each polynomial,
up to some universal scaling factor. We can express this $d$-degree polynomial $\hat{f}_i$ via Lagrange interpolation:
\begin{align*}
    L_{d, j}(x) =&\;  \prod_{k \neq j}^{d} \frac{x - x_k}{x_j - x_k} ;\\
    \hat{f}_i(x) =&\; \sum_{j=0}^d \hat{y}_j L_{n,j}.
\end{align*}
Note that $\sum_i \hat{f}_i(v_U) = 1$ as the scores coincide exactly with our query results at the uniform vector. 
To analyze the representation error, let $\{f^*_i\}$ be the set of true polynomials $f_i$ rescaled to sum to 1 at the uniform vector; this involves dividing by a factor $S \in [n\lambda, n]$, and produces identical scores at every point. Consider the difference $\abs{ \hat{y}_j - y^*_j }$ for each $y^*_j = f^*_i(x_j)$. The query error for $\hat{y}_j$ prior to rescaling is at most $\beta$; rescaling by $R^*_v$ would increase this to at most $3\beta / \lambda$, which is amplified to at most 
\begin{align*}
    \abs{ \hat{y}_j - y^*_j } \leq&\; \frac{3 \beta }{\lambda} + \frac{2n\beta}{\lambda} \leq \frac{3n\beta}{\lambda}
\end{align*}
as each query score is at most 1 (and our setting is trivial for $n\leq 2$).
The magnitude of each of the $d+1$ Lagrange terms can be bounded by:
\begin{align*}
   \abs{ L_{d,j}(x) } \leq&\; \prod_{j=1}^{d/2} \frac{Z^2}{j^2}  \\
   \leq&\; \frac{Z^d}{((d/2)!)^2}
\end{align*} 
for any $x \in [0, 1]$, and so for any function $\hat{f}_i(x)$ we can bound its distance from $f^*_i(x)$ by:
\begin{align*}
    \abs{f^*_i(x) - \hat{f}_i(x)} =&\; (d+1) \cdot \frac{3n\beta Z^d}{\lambda ((d/2)!)^2 } \\ 
   \leq&\; \frac{(d+1)3n\beta Z^d}{\lambda 2^{d/2}}. \\
\end{align*}
This holds simultaneously for each $\hat{f}_i$ which, using the fact that the true ratio is at least $\lambda/n$ and the per-function bound applies to each denominator term, gives us a total bound on the score estimates we generate:
\begin{align*}
   \abs{ \frac{\hat{f}_i(x)}{\sum_{j=1}^x \hat{f}_j(x)}  - \frac{ f_i(x)}{\sum_{j=1} f_j(x)}} \leq&\; \parens{ 1 + \frac{(d+1)3n\beta Z^d}{\lambda 2^{d/2}} } \cdot \frac{(d+1)3n^3\beta Z^d }{\lambda^2 2^{d/2} } \\
   \leq&\; \frac{7n^3d\beta Z^d}{\lambda^2 2^{d/2} } \\
   \leq&\; \frac{3\cdot (6nd)^{d/2 + 2} \beta }{\alpha^d \lambda^2 2^{d/2} } \\
   =&\; \frac{(3nd)^{d/2 + 2} \beta}{\alpha^d \lambda^2}.
\end{align*}
Taking $\beta \leq \frac{\epsilon \alpha^d \lambda^2 }{(3nd)^{d/2 + 2}}$ gives us an absolute error of at most $\epsilon$ per item score, satisfying a Euclidean bound of $\epsilon$ from any true score vector $M(w)/M^*_w$ for our hypothesis $\hat{M}(v) = \{\hat{f}_i(v_i): i \in [n]\}$.
\end{proof}

\subsection{Proofs of Multivariate Polynomial Local Learnability}

Recall that the two classes of multivariate polynomial models we consider are {\it bounded-degree multilinear polynomial} preference models $\M_{BMLP}$, where:
\begin{itemize}
    \item for each $i$, $M(v)_i = f_i(v)$, where $f_i$ is a degree-$d$ multilinear (i.e.\ linear in each item) polynomial which takes values in $[\lambda, 1]$ over $\Delta(n)$  for some constant $\lambda > 0$, 
\end{itemize}
and the class of 
{\it bounded-degree normalized multivariate polynomial} preference models $\M_{BNMP}$, where:
\begin{itemize}
    \item for each $i$, $M(v)_i = f_i(v)$, where $f_i$ is a degree-$d$ polynomial which takes values in $[\lambda, 1]$ over $\Delta(n)$  for some constant $\lambda > 0$, where $\sum_i f_i(v) = C$ for some constant $C$. 
\end{itemize}

We prove local learnability results for each case.

\begin{lemma}
$\M_{BMLP}$ is 
$O(n^d)$-locally learnable by an algorithm $\A_{BMLP}$ with $\beta \leq O(\frac{\epsilon^2 }{ \textup{poly}(n (d / \alpha)^d)} )$. 
\end{lemma}

\begin{proof}
Consider the set of polynomials where each $v_n$ term is reparameterized as $1 - \sum_{i=1}^{n-1} v_i$, then translated so that the uniform vector appears at the origin (i.e.\ with $x_n = -\sum_{i=1}^{n-1} x_i$). Our approach will be to learn a representation of each polynomial normalized their sum, which is unique up to a universal scaling factor. Let $f_i^*$ be the representation of $f_i$ in this translation. Consider the $N = \sum_{j=0}^d{n - 1 \choose j}$-dimensional basis $\mathcal{B}$ where each variable in a vector $x$ corresponds to a monomial of at most $d$ variables in $v$, each with degree 1, with the domain constrained to ensure mutual consistency between monomials, e.g.:
\begin{align*}
    \mathcal{B} =&\; \{ 1, v_1, \ldots, v_{n-1}, v_1 v_2, \ldots, \prod_{j=n-d}^{n-1} v_{j} \}. 
\end{align*}
Observe that each $f_i^*$ is a linear function in this basis.
Let $q_{i}(x) = M(v)_i / M^*_v$ denote the normalized score for item $i$ at $v$, where $v$ translates to $x$ in the new basis. For we any $x$ we have:
\begin{align*}
    \frac{f_i^*(x)}{\sum_{j=1}^n f_j^*(x)} = q_{i}(x),
\end{align*}
and let $\hat{q}_i(x)$ denote the analogous perturbed query result, both of which sum to 1 over each $i$.  We are done if we can estimate the vector $q(x)$ up to distance $\epsilon$ for any $x$.

With $f^*_i(x) = \langle a, x \rangle + a_0$ and $\sum_{i=1}^{n} f_i^*(x) = \langle b, x \rangle + b_0$, our strategy will be to estimate the ratio of each coefficient with $b_0$, for each $f^*_i$, in increasing order of degree. While our parameterization does not include item $n$, we will explicitly estimate $b$ separate from each $a$, which we can then use to estimate $f_n^*(x) = \langle b, x \rangle + b_0 - \sum_{i=1}^{n-1} f^*_i(x)$. For a monomial $m$ of degree $j$, we can estimate its coefficient for all $f^*_i$ simultaneously by moving the values for variables it contains simultaneously from the $\mathbf{0}$ vector, and viewing the restriction to its subset monomials as a univariate polynomial of degree $j$. We will use a single query to the $\mathbf{0}$ vector, and $2j+1$ additional queries for each degree-$j$ monomial (which can be used for learning that monomial's coefficient in all $f^*_i$ simultaneously), resulting in a total query count of:
\begin{align*}
    1 + \sum_{j=1}^d (2j+1) \cdot {n-1 \choose j} =&\; 1 + \sum_{j=1}^d (2j+1)  \frac{(n-1)!}{j! (n-j-1)!} \\
    =&\; O(n^d).
\end{align*}
Querying $\mathbf{0}$ gives us an estimate for each additive term:
\begin{align*}
    \frac{\hat{a}_0^i }{b_0} =&\; \hat{q}_i(\mathbf{0})
\end{align*}
which sum to 1 over all items (and we will take $\hat{b}_0 = 1$). We now describe our strategy for computing higher-order coefficients in terms of lower-order coefficients under the assumption of {\it exact} queries, after which we conduct error propagation analysis.
For a monomial $m$ of degree $j$, let $x_{(h,m)}$ be the point where $x_{(h,m),i} = hZ$ if an item $i$ belongs to $m$ and $0$ otherwise, with higher degree terms satisfying the basis constraints (i.e. $(hZ)^{3}$ for a degree-3 subset of $m$, and $(hZ)^j$ for $m$), which also results in the term for a monomial containing any item not in $m$ being set to zero. Query $x_{(h,m)}$ for $2j + 1$ distinct values $h$ in $\{\pm 1, \ldots, \pm (j+1)\}$. For $Z = \alpha/(2d(d+1))$ all queries lie in the $\alpha$-ball, as the $\ell_1$ norm of the positive coefficients, as well as the negative offset for item $n$, are both bounded by $\alpha/2$ in the original simplex basis.
Suppose all coefficients up to degree $j-1$ are known. The result of such a query (with $z = hZ$) is equivalent to:
\begin{align*}
    q(x_{(h,m)}) =&\; \frac{a_m z^j + f_a(z)}{b_m z^j + f_b(z)}
\end{align*}

where $f_a$ and $f_b$ are $(j-1)$-degree univariate polynomials, where each coefficient of some degree $k \leq j-1$ is expressed by summing the coefficients for degree-$k$ monomials which are subsets of $m$, for $a$ and $b$ respectively. Rearranging, we have:
\begin{align*}
    a_m =&\; q_i(x_{(h,m)}) \cdot b_m + \frac{q(x_{(h,m)})\cdot f_b(z) - f_a(z)}{z^j}.
\end{align*}
This gives us a linear relationship between $a_m$ and $b_m$ in terms of known quantities after just one query where $z \neq 0$.
Suppose we could make {\it exact} queries; if we observe two distinct linear relationships, we can solve for $a_m$ and $b_m$. If each query gives us the same linear relationship, i.e. $q_i(x_{(h,m)}) = q_i(x_{(h',m)})$ for every query pair $(h, h')$, then equality also holds for each of the $(q_i(x_{(h,m)})\cdot f_b(z) - f_a(z)) / {z^j}$ terms. If the latter term is truly a constant function $c$:
\begin{align*}
    \frac{q_i(x_{(h,m)})\cdot f_b(z) - f_a(z)}{z^j} = c
\end{align*}
then we also have:
\begin{align*}
  {(a_m z^j + f_a(z)) \cdot f_b(z) - (b_m z^j + f_b(z)) \cdot f_a(z)} =&\; c z^j (b_m z^j + f_b(z)).
\end{align*}
Each side is a polynomial with degree at most $2j$, and thus cannot agree on $2j + 1$ points unless equality holds. However, if equality does hold, we have that either $c=0$ or $b_m  = 0$, as the left side has degree at most $2j-1$, and both $z^j$ and $b_m z^j + f_b(z)$ are bounded away from 0 for any $z \neq 0$. If $c \neq 0$, then we have that $b_m = 0$ and $a_m = c$.
If $c=0$, then we have 
\begin{align*}
    a_m z^j f_a(z) f_b(z) -  b_m z^j f_a(z) f_b(z) =&\; 0,
\end{align*}
which implies $a_m = b_m$, as $f_a(z) f_b(z)$ cannot be equal to 0 everywhere due to each $a^i_0$ and $b_0$ being positive. Our answer to $q(x_{(h,m)})$ will be bounded above 0 and below 1, allowing for us to solve for both $a_m$ and $b_m$ as 
\begin{align*}
    a_m = b_m =&\; \frac{q_i(x_{(h,m)})\cdot f_b(z) - f_a(z)}{(1 - q_i(x_{(h,m)})) z^j}.
\end{align*}
To summarize, if given exact query answers for $2j+1$ distinct points, we must be in one of the following cases:
\begin{itemize}
    \item We observe at least two distinct linear relationships between $a_m$ and $b_m$ from differing query answers;
    \item We observe a non-zero constant $\frac{q_i(x_{(h,m)})\cdot f_b(z) - f_a(z)}{z^j} = c$ for each query, and have $a_m = c$;
    \item We observe $\frac{q_i(x_{(h,m)})\cdot f_b(z) - f_a(z)}{z^j} = 0$ for each query, and can solve for $a_m = b_m$.
\end{itemize}

To begin our error analysis for perturbed queries, we first show a bound on the size of the coefficients for a polynomial which is bounded over a range.

\begin{lemma}
Each degree-$d'$ coefficient of $f_i^*$ is at most ${d'}^{2d'}$.
\end{lemma}
\begin{proof}
First note that the constant coefficient and the coefficient for each linear term have magnitude at most 1, as the function is bounded in $[\lambda, 1]$ over the domain (which includes $\mathbf{0}$).
For a degree-$d'$ monomial $m$, consider the univariate polynomial corresponding to moving each of its variables in synchrony while holding the remaining variables at 0, whose degree-$d'$ coefficient is equal to $a_m$. 
Consider the Lagrange polynomial representation of this polynomial 
\begin{align*}
    L_{d', j}(x) =&\;  \prod_{k \neq j}^{d'} \frac{x - x_k}{x_j - x_k} ;\\
    \hat{f}_i(x) =&\; \sum_{j=0}^{d'} \hat{y}_j L_{n,j}.
\end{align*}
for $d'+ 1$ evenly spaced points in the range $[-1/n, 1/d' - 1/n]$, which are all feasible under the simplex constraints (corresponding to $v_i \in [0, 1/d']$ in the original basis, for each $i \in m$). Each pair of points is separated by a distance of at least $1 / ({d'}^2)$, and so the leading coefficient of each Lagrange term is at most $ {d'}^{2(d'-1)}$. Each $\hat{y}_j$ is in $[\lambda, 1]$ and so we have
\begin{align*}
    a_m \leq&\; (d' + 1) {d'}^{2(d'-1)} \\
    \leq&\; {d'}^{2d'}
\end{align*}
for each $d' > 1$.
\end{proof}

As we estimate coefficients for monomials of increasing degree, we will maintain the invariant that each degree-$j$ coefficient of $a$ and $b$ is estimated up to additive error $\epsilon_j$, with respect to the normalization where $b_0 = 1$. Immediately we have $\epsilon_0 = \beta$ for the estimates $\hat{a}_0$ from our query to the $\mathbf{0}$ vector. We will also let $\beta_j$ denote the error of a polynomial $\hat{f}_a$ restricted to terms for subsets of a $j$-degree monomial $m$

For a monomial $m$, suppose we receive 2 queries $\hat{q}_{i}(x_{(h, m)})$ and $\hat{q}_{i}(x_{(h', m)})$ for some $h$ and $h'$ where
\begin{align*}
    \abs{ \hat{q}_{i}(x_{(h, m)}) - \hat{q}_{i}(x_{(h', m)}) } \geq F_j
\end{align*}
for some quantity $F_j$. 
Then we have:

\begin{align*}
    \hat{a}_m  =&\; \hat{q}_{i}(x_{(h, m)})  \hat{b}_m  +  \frac{\hat{q}_{i}(x_{(h, m)}) \cdot \hat{f}_b(hZ) - \hat{f}_a(h Z)}{(hZ)^j }   \\
    =&\; \hat{q}_{i}(x_{(h', m)})  \hat{b}_m  +  \frac{\hat{q}_{i}(x_{(h,' m)}) \cdot \hat{f}_b(h' Z) - \hat{f}_a(h' Z)}{(h' Z)^j }   \\
    \hat{b}_m =&\; \frac{\hat{a}_m}{\hat{q}_{i}(x_{(h, m)})} +  \frac{ \frac{ \hat{f}_a(h Z) }{\hat{q}_{i}(x_{(h, m)})} - \hat{f}_b(hZ)}{(hZ)^j } ; \\
    =&\; \frac{\hat{a}_m}{\hat{q}_{i}(x_{(h', m)})} +  \frac{ \frac{ \hat{f}_a(h' Z) }{\hat{q}_{i}(x_{(h', m)})} - \hat{f}_b(h'Z)}{(h'Z)^j } ; \\
    \frac{\hat{a}_m}{\hat{q}_{i}(x_{(h', m)})}  - \frac{\hat{a}_m}{\hat{q}_{i}(x_{(h, m)})} =&\;   \frac{ \frac{ \hat{f}_a(h Z) }{\hat{q}_{i}(x_{(h, m)})} - \hat{f}_b(hZ)}{(hZ)^j } - \frac{ \frac{ \hat{f}_a(h' Z) }{\hat{q}_{i}(x_{(h', m)})} - \hat{f}_b(h'Z)}{(h'Z)^j } ; \\
    \hat{a}_m  =&\;   \frac{ \frac{\hat{q}_{i}(x_{(h', m)}) \hat{f}_a(h Z) }{\hat{q}_{i}(x_{(h, m)})} - \hat{q}_{i}(x_{(h', m)}) \hat{f}_b(hZ)}{ \parens{ 1  - \frac{\hat{q}_{i}(x_{(h', m)})}{\hat{q}_{i}(x_{(h, m)})} } \cdot (hZ)^j } - \frac{  \hat{f}_a(h' Z)  - \frac{ \hat{f}_b(h'Z)}{\hat{q}_{i}(x_{(h', m)})} }{ \parens{ 1  - \frac{\hat{q}_{i}(x_{(h', m)})}{\hat{q}_{i}(x_{(h, m)})} } \cdot (h'Z)^j } ; \\
    \hat{b}_m =&\; \frac{ \frac{\hat{q}_{i}(x_{(h, m)}) \cdot \hat{f}_b(hZ) - \hat{f}_a(h Z)}{(hZ)^j } - \frac{\hat{q}_{i}(x_{(h,' m)}) \cdot \hat{f}_b(h' Z) - \hat{f}_a(h' Z)}{(h' Z)^j } }{\hat{q}_{i}(x_{(h', m)}) - \hat{q}_{i}(x_{(h', m)})};
\end{align*}    
where $\hat{f}_a$ and $\hat{f}_b$ are the univariate polynomials from summing the lower-order coefficient estimates for each degree up to $j-1$.
The additive error to each $\hat{f}_a(hZ)$ and $\hat{f}_a(hZ)$ can be bounded by:
\begin{align*}
    \beta + \sum_{k=1}^{j-1} {n \choose k} (hZ)^k k^{2k} \epsilon_k =&\;\beta + \sum_{k=1}^{j-1} {n-1 \choose k} (k^2hZ)^k \epsilon_k. \\
\end{align*}
Further, the magnitude of each $\hat{f}_a(hZ)$ and $\hat{f}_b(hZ)$ is at most $1 + \sum_{k=1}^{j-1} {n -1 \choose k} (k^2hZ)^k$.
We can bound the error of other terms as follows:
\begin{itemize}
    \item Each $\hat{q}_{i}(x_{(h', m)}) - \hat{q}_{i}(x_{(h', m)})$ has magnitude at least $F_j$ and at most 1, and additive error at most $2\beta$;
    \item Each $\hat{q}_{i}(x_{(h', m)})$ has value at least $\frac{\lambda}{n}$ and at most 1, and additive error at most $\beta$;
    \item Each $\frac{\hat{q}_{i}(x_{(h', m)})}{\hat{q}_{i}(x_{(h, m)})}$ term is either greater than $\frac{1}{1 - F_j}$ or at most $1 - F_j$; the true ratio between the numerator and denominator is at least $\lambda / n$ most $n / \lambda$, with additive error up to $\beta$ in both.
    \item Each $1 - \frac{\hat{q}_{i}(x_{(h', m)})}{\hat{q}_{i}(x_{(h, m)})}$ term, is either greater than $F_j$ or at most $1 - \frac{1}{1 - F_j}$;
    \item Each $(hZ)^j$ has magnitude at least $Z^j$;
\end{itemize}

The error in the numerator of $\hat{a}_m$, and the fractional terms in the numerator of $\hat{b}_m$ is dominated by multiplying the functions of $\hat{q}_i$ with the polynomials themselves. As such, we can bound the error to $a_m$ and $b_m$ by $\epsilon_j$ if we have that:
\begin{align*}
    \epsilon_j \geq&\; O\parens{   \frac{n \beta  }{ \lambda F_j Z^j } \cdot \parens{ 1 + \sum_{k=1}^{j-1} {n -1 \choose k} (k^2hZ)^k } } \\
    =&\; O\parens{   \frac{n \beta  }{ \lambda F_j Z^j } \cdot \parens{ 1 + \sum_{k=1}^{j-1} {n-1  \choose k} (h \alpha)^k } } \\
    =&\; O\parens{ \frac{nd^{2j} \beta }{\lambda \alpha^j F_j }  }
\end{align*}
for any $\alpha < 1/(nd)$.
Now suppose all pairs of query answers we see are separated by less than $F_j$; the additive error to each estimate of the quantity
\begin{align*}
    \hat{c}_{(h,m)} =&\; \frac{\hat{q}_i(x_{(h,m)})\cdot \hat{f}_b(z) - f_a(z)}{(hZ)^j}
\end{align*}
is $\mathcal{E}_j = O\parens{ \frac{\beta}{Z^j} \cdot \parens{ 1 + \sum_{k=1}^{j-1} {n \choose k} (k^2hZ)^k }} = O\parens{\beta \cdot nd^{2j}/\alpha^j}$. If each such quantity has value at most $\mathcal{E}_j$, we assume this quantity is zero and solve for $a_m = b_m$. 
If some are larger, we must be in the case where $\hat{b}_m \approx 0$
and so we set $a_m = \hat{c}_{(h,m)}$ for any query result.
By taking each $F_j = O(\sqrt{\beta}\poly(n, d^j, 1/\alpha^j))$ we can obtain a bound of $\epsilon_j = O(\sqrt{\beta} \poly(n, d^j, 1/\alpha^j))$ to each coefficient regardless of which case we are in; after summing the error contribution across coefficients and accounting for renormalization, recalling that $\lambda = \Omega(1/n)$, we obtain a bound of $\epsilon$ on score vector errors (for any desired norm) provided that $\epsilon \geq \sqrt{\beta}\poly(n, d^d, 1/\alpha^d)$.
\end{proof}

Next, we prove the local learnability result for normalized multivariate polynomials.

\begin{lemma}
$\M_{BMNP}$ is 
$O(n^d)$-locally learnable by an algorithm $\A_{BMNP}$ with $\beta \leq \frac{\epsilon}{\alpha^{d} F(n, d)} $, where $F(n,d)$ is some function depending only on $n$ and $d$ which is finite for all $n,d \in \mathbb{Z}$. 
\end{lemma}

\begin{proof}
Our approach will be to construct a set of $O(n^d)$ queries which results in a data matrix which is nonsingular in the space of $d$-degree multivariate polynomials, solve for the coefficients of each $f_i$ as a linear function over this basis, and show that the basis is sufficiently well-conditioned such that our approximation error is bounded.

Consider the set of polynomials where each $v_n$ term is reparameterized as $1 - \sum_{i=1}^{n-1} v_i$, then translated so that the uniform vector appears at the origin (i.e.\ with $v_n = -\sum_{i=1}^{n-1} v_i$). Our approach will be to learn a representation of each polynomial directly, as they are already normalized to sum to a constant (which must be in the range $[1,n]$). Let $f_i^*$ be the representation of $f_i$ in this translation. Let $\mathcal{B}$ be the 
$N = \sum_{j=0}^d (n-1)^j$-dimensional 
basis where each variable in a vector $x$ corresponds to a monomial of variables in $v$ with degree at most $d$, with the domain constrained to ensure mutual consistency between monomials, e.g.:
\begin{align*}
    \mathcal{\B} =&\; \{ 1, v_1, \ldots, v_{n-1}, v_1^2, v_1 v_2, \ldots, v_{n-1}^d \}. 
\end{align*}
Observe that $f_i^*$ is a linear function in this basis, with $f^*_i(x) = \langle a, x \rangle $ and $\sum_{i=1}^n f_i^*(x) = \langle b, x \rangle $ for any $x$ represented in $\mathcal{B}$.  

There is a large literature on constructing explicit query sets for multivariate polynomial interpolation, which ensure that the resulting data matrix is nonsingular; see \cite{Gasca2000} for an overview. The set must have at least $N$ points to ensure uniqueness of interpolation, and this is sufficient when points are appropriately chosen. Let $S^*$ be any such set such that each point $\norm{w}_1 \leq 1/2$ for each $w \in S^*$, and let $C_{n,d}$ be the $\ell_{\infty}$ condition number of the resulting matrix $Y$ (which will be positive due to nonsingularity) given by:
\begin{align*}
Y =&\;
\begin{bmatrix}
y^{(1)}_1 &  \cdots & y^{(1)}_N \\
\vdots    &               &      \vdots  \\
y^{(j)}_1 & \cdots & y^{(j)}_N  \\
\vdots    &            &      \vdots  \\
y^{(N)}_1 & \cdots & y^{(N)}_N \\
\end{bmatrix}
\end{align*}
where $y^{(j)}$ is the representation of $s^{(j)}$ in the basis $\mathcal{B}$. We show that for any $\alpha$, we can construct a matrix $X$ from a query set $S^{\alpha}$ of size $N$ where  $\norm{v}_1 \leq \alpha/2$ for each $v \in S^{\alpha}$. For each $s^{(j)}$, let $v^{(j)} = \alpha s^{(j)}$, which results in $\norm{v}_1 \leq \alpha/2$ for the parameterization over $n-1$ items, and so radius of $\alpha$ holds when including all $n$ items. This results in a matrix $X$ given by
\begin{align*}
X =&\;
\begin{bmatrix}
x^{(1)}_1 &  \cdots & x^{(1)}_N \\
\vdots    &               &      \vdots  \\
x^{(j)}_1 & \cdots & x^{(j)}_N  \\
\vdots    &            &      \vdots  \\
x^{(N)}_1 & \cdots & x^{(N)}_N \\
\end{bmatrix}
\end{align*}
We then have
\begin{align*}
   X =&\; YD,
\end{align*}
where $D$ is a diagonal matrix with the $j$th diagonal entry $\nu_j$ equal to $\alpha^{d_j}$, where $d_j$ is degree of the $j$th monomial in $\mathcal{B}$, as our scaling by $\alpha$ is amplified for each column in correspondence with the associated degree; the values of $D$ will range from $\alpha^d$ to 1. We can then bound the condition number of $X$ as:
\begin{align*}
    \textup{cond}(X) =&\; \textup{cond}(YD) \\
    =&\; \norm{YD}\norm{(DY)^{-1}} \\
    \leq&\; \norm{Y}\norm{D}\norm{D^{-1}}\norm{Y^{-1}} \\
    =&\; \textup{cond}(Y) \cdot \textup{cond}(D) \\
    \leq&\; C_{n,d} \frac{\max_j \nu_j}{\min_j \nu_k} \\
    =&\; \frac{C_{n,d} }{\alpha^d}.
\end{align*}

Let $q$ denote the vector of exact answers to each query in $x$ from $f_i$, equal to $a \dot x$ and let $\hat{q}$ be the answers we observe for item $i$ from querying each $x$. As $X$ is nonstationary, we have that $Xa = q$, and by standard results in perturbation theory for linear systems, for $\hat{a}$ such that $X\hat{a} = \hat{q}$ we have that:
\begin{align*}
    \frac{\norm{ \hat{a} - a}}{\norm{a}} \leq&\; \textup{cond}(X) \frac{\norm{ \hat{q} - q} }{\norm{q}} \\
    \leq&\; \frac{\beta n C_{n,d} }{k^2 \alpha^d}
\end{align*}
as each entry in $q$ is at least $\lambda \geq k^2/n$. Further note that the maximum coefficient of a degree-$d$ multivariate polynomial which takes maximum value 1 over the unit ball (and hence the simplex) can be shown to be bounded by a finite function of $n$ and $d$ (see \cite{Kellogg1928}); when accounting for this factor in relative error across all terms and items, as well as the condition number, we have that for $\beta \leq \frac{\epsilon}{\alpha^d F(n,d)}$ for some function $F(n,d)$, the scores generated by the functions $\hat{f}_i$ using our estimated coefficients $\hat{a}$ result in score vector estimates bounded by $\epsilon$.

\end{proof}

\newcommand{\halpha}{\hat{\alpha}}
\newcommand{\cN}{\mathcal{N}}
\newcommand{\vu}{v_{\textnormal{unif}}}
\newcommand{\hs}{\hat{s}}
\newcommand{\hf}{\hat{f}}
\newcommand{\hy}{\hat{y}}
\newcommand{\Ot}{\widetilde{O}}

\subsection{Proof of SFR Local Learnability}
We now prove that functions with local sparse Fourier transformation are locally learnable.
Recall that a function $f(x)$ has a $\ell$-sparse
Fourier transform if it can be written
as 
\[
f(x) = \sum_{i=1}^\ell \xi_i e^{2\pi \mathbf{i} \eta_i x}
    \,,
\]
where $\eta_i$ is the $i$-th frequency, $\xi_i$ is the corresponding magnitude, and $\mathbf{i} = \sqrt{-1}$.

We will use the following result 
about learning sparse Fourier transforms \cite{PriceS15}.

\begin{theorem}[\cite{PriceS15}]
\label{thm:learn_fourier}
Consider any function $f(x) : \R \rightarrow \R$
of the form 
\[
f(x) = f^*(x) + g(x)
    \,,
\]
where $f^*(x) = \sum_{i=1}^\ell \xi_i e^{2\pi \mathbf{i} \eta_i x}$ with frequencies $\eta_i \in [-F,F]$ and 
frequency separation $\hat{\alpha} = \min_{i \neq j}|\eta_i - \eta_j|$, and
$g(x)$ is the arbitrary noise function.
For some parameter $\delta > 0$, we define the \emph{noise-level} over an interval $I = [a,b] \subseteq \R$
as
\[
\mathcal{N}^2 = \frac{1}{|I|} \int_{I} |g(x)|^2 dx + \delta \sum_{i=1}^\ell|\xi_i|^2
    \,.
\]
There exists an algorithm that takes samples 
from the interval $I$ with length $|I| > O(\frac{\log(\ell/ \delta)}{ \hat{\alpha}})$ and returns a set of $\ell$ pairs
$\{(\xi_i', \eta_i')\}$ such that 
for any $|\xi_i| = \Omega(\cN)$ we have for an 
appropriate permutation of the indices
\[
|\eta_i - \eta_i'| = O\Big(\frac{\cN}{ |I| |\xi_i|}\Big),
\qquad | \xi_i - \xi_i'|  = O(\cN), \forall i \in [\ell]
    \,.
\]
The algorithm takes $O(\ell \log(F |I|) \log(\frac{\ell}{\delta}) \log(\ell))$ samples
and 
$O(\ell \log(F |I|) \log(\frac{F|I|}{\delta}) \log(\ell))$
and succeeds with probability at least $1-1/k^c$
for any arbitrary constant $c$.
\end{theorem}
Furthermore, the algorithm used in the above theorem
uses samples of the form $x_0, x_0 + \sigma \cdots x_0 + \ell \log(\ell/\delta) \sigma$
for randomly chosen $x_0$ and $\sigma = O(|I|/\ell \log(\ell/\delta))$.

We will use the above theorem to learn the sparse
Fourier representation of the preference model.
Recall that for a memory vector $v$ and item $i \in [n]$,
$M(v)_i = f_i(v_i)$.

\begin{proof}
Let $\vu$ denote the uniform memory vector.
We will learn each function $f_i$ separately.
Fix $i \in [n]$.
We will set the interval $I$ to be $[1/n-Z, 1/n + Z]$
for some sufficiently small $\frac{\log(\ell/\delta)}{\halpha} \leq  Z \leq \alpha/2$
where $\halpha$ is the frequency separation, where $\alpha = \Tilde{\Omega}(1/\hat{\alpha})$ so that $Z$ is defined. 
Let $S = \{ x_j\}_{j = 1}^{\tilde{O}(\ell)}$
for $x_j \in [-Z, Z]$
be a set of points
such that the Fourier learning algorithm queries $1/n +x$ for each $x \in S$.
For each point $x \in S$, we define the
memory vector $v^x = \vu + x e_i - x e_j$ 
where $j$ is a fixed randomly chosen other index.
All such vectors lie in $V_{\alpha}$, as $2 (\alpha/2)^2 \leq \alpha^2$.
We query all vectors $v^x$ for $x \in S$, along with $\vu$.
Recall that $\hs_v$ is the empirical score vector
at a memory vector $v$.
For each vector $v$,
let $R_v$ be the sum of all scores of all the $n-2$ items held at $\frac{1}{n}$, divided by the sum of those same items' scores in the uniform vector $\vu$. 
For each vector $v^x$ we multiply the score $\hs_{v^x,i}$ of item  $i$ by $R_{v^x}$
to obtain  a noisy sample of $f_i(1/n+x)$. 
For $i \in \Ot(\ell)$, 
let the $i$-th sample be denoted by $\hy_i$
and the true value $f_i(1/n+x_i)$ be denoted by $y_i$.
We then pass all these samples to the Fourier learning algorithm  in Theorem~\ref{thm:learn_fourier} in order to get an estimate $\hf$
of $f$.

We now analyze the error in the samples.
Let $R_v^*$ be the corresponding ratio of these sums of scores under $\{ f_i \}$; each sum is within $[\frac{\lambda}{3}, 1]$ at each vector, and the sums of observed scores have additive error at most $2n\beta$. 
As such, $R_v$ has additive error at most $\frac{2n\beta}{\lambda}$ from $R_v^*$.
For each vector $v^x$ we have that $\hs_{v^x, i}/(\sum_j \hs_{v^x, j})$ is within a $\beta$ error from $s_{v^x, i}/(\sum_j s_{v^x, j})$.
Hence, the total error in each sample is bounded as:
\[
\abs{\hy_i - y_i} \leq \frac{7n \beta}{\lambda}
    \,.
\]
Using this we can bound the total noise term by 
$\cN = 8n \beta/\lambda $
using our choice of $\delta = (\beta n)/(\lambda \sum_{i=1}^\ell|\xi_i|)$.
The algorithm will return a set of $\{(\hat{\eta}_i, \hat{\xi}_i)\}$ such that 
\[
|\eta_i - \eta_i'| = O\Big(\frac{1}{\alpha}\Big),
\qquad | \xi_i - \xi_i'|  = O(\frac{\beta n}{\lambda}), \forall i \in [\ell]
    \,.
\]
So for function $\hat{f}_i(x)$ we can bound its distance from $f_i(x)$ by:
\begin{align*}
    \abs{f_i(x) - \hat{f}_i(x)} =&\; 
    \abs{\sum_{i=1}^\ell \xi_i e^{2\pi \mathbf{i} \eta_i x} - \sum_{i \in [\ell]} \hat{\xi}_i e^{2\pi \mathbf{i} \hat{\eta}_i x}} \\
    \leq &\; \sum_{i \in [\ell]} \abs{ \xi_i e^{2\pi \mathbf{i} \eta_i x} - \hat{\xi}_i e^{2\pi \mathbf{i} \hat{\eta}_i x}} \\
    \leq &\; \sum_{i \in [\ell]} \abs{ \xi_i - \hat{\xi}_i } \abs{\eta_i - \hat{\eta}_i} \\
   \leq&\; O(\frac{\ell n\beta}{\lambda \alpha}), \\
\end{align*}
since we normalize the above estimates to get a score estimate, the total bound on the score estimates can be bounded as: 
\begin{align*}
   \abs{ \frac{\hat{f}_i(x)}{\sum_{j=1}^x \hat{f}_j(x)}  - \frac{ f_i(x)}{\sum_{j=1} f_j(x)}} \leq&\; O(\frac{\ell\beta n}{\alpha\lambda}).
\end{align*}
Taking $\beta \leq \frac{\epsilon \lambda \alpha }{ \sqrt{n} \ell}$ gives us an error of at most $\epsilon \sqrt{n}$, satisfying a Euclidean bound of $\epsilon$ from any true score vector $M(w)/M^*_w$ for our hypothesis model $\hat{M}(v) = \{\hat{f}_i(v_i): i \in [n]\}$.
\end{proof}

\section{Omitted Proofs for Section \ref{sec:contracting}}
\label{sec:rcfkm-proofs}

\subsection{Proof of Theorem \ref{alg:bandit-constricting}}

\begin{proof}
First observe that $y_t \in \K_t$ every round, as 

For $x^* = \textup{arg min}_{x \in \K_T} \sum_{t=1}^T f_t(x)$, let $x^*_{\delta,\epsilon} = \Pi_{\K_{T, \delta,\epsilon}}(x^*)$. By linearity and properties of projection, we also have that $x^*_{\delta,\epsilon} = \textup{arg min}_{x \in \K_{T,\delta, \epsilon}} \sum_{t=1}^T f_t(x)$, and that $\norm{x^*_{\delta,\epsilon} - x^*} \leq (\delta + \epsilon)\frac{D}{r}$. For $G$-Lipschitz losses $\{f_t\}$ we have
\begin{align*} 
\sum_{t=1}^T \E[\phi_t] - \sum_{t=1}^T f_t(x^*) =&\;  \sum_{t=1}^T \E[f_t(y_t)] - \sum_{t=1}^T f_t(x^*) \\
    \leq&\; \sum_{t=1}^T \E[f_t(y_t)] - \sum_{t=1}^T f_t(x^*_{\delta, \epsilon}) + \delta TG\frac{D}{r} + \epsilon TG\frac{D}{r}.
\end{align*}
Let $\hat{f}_t(x)  = \E_{u \sim \mathbb{B}}[f(x + \delta u + \xi_t)] = f_t(x + \xi_t) $ by linearity. Then we can bound the regret by:
\begin{align*}
 \sum_{t=1}^T \E[\phi_t] - \sum_{t=1}^T f_t(x^*) \leq&\;  \sum_{t=1}^T \E[f_t(y_t)] - \sum_{t=1}^T f_t(x^*_{\delta,\epsilon})  + \frac{\delta TGD}{r} + \frac{\epsilon TGD}{r}\\
 =&\;  \sum_{t=1}^T \E[\hat{f}_t(x_t)] - \sum_{t=1}^T f_t(x^*_{\delta,\epsilon})  +  \frac{\delta TGD}{r} +  \frac{\epsilon TGD}{r} \\
  \leq&\; \sum_{t=1}^T \E[\hat{f}_t(x_t)] - \sum_{t=1}^T \hat{f}_t(x^*_{\delta,\epsilon}) + \frac{\delta TGD}{r} + \epsilon TG\parens{\frac{D}{r} +1} \\
    \leq&\; \sum_{t=1}^T \E[\hat{f}_t(x_t)] - \sum_{t=1}^T \hat{f}_t(x^*_{\delta,\epsilon}) + \frac{\delta TGD}{r} +  \frac{2 \epsilon TGD}{r}
 \end{align*} 
 
Next, we prove a series of lemmas --- an analysis of online gradient descent for contracting decision sets, and a corresponding bandit-to-full-information reduction --- which allow us to view the remaining summation terms involving $\{x_t\}$ as the expected regret of stochastic online gradient descent for the loss function sequence $\{\hat{f}_t\}$ with respect to $\K_{T, \delta, \epsilon}$.

When modifying online gradient descent to project into smaller sets each round, the analysis is essentially unchanged.

\begin{algorithm}
\caption{Contracting Online Gradient Descent.}\label{alg:ogd-contracting}
\begin{algorithmic}
\State Input: sequence of contracting convex decision sets $\K_1, \ldots \K_T$, $x_1 \in \K_1$, step size $\eta$
\State Set $x_1 = \mathbf{0}$
\For{$t = 1$ to $T$}
    \State Play $x_t$ and observe cost $f_t(x_t)$
    \State Update and project: \begin{align*}
        y_{t+1} =&\; x_t - \eta \nabla \ell_t(x_t) \\
        x_{t+1} =&\; \Pi_{\K_{t+1}}(y_{t+1})
    \end{align*}
\EndFor
\end{algorithmic}
\end{algorithm}

\begin{lemma} \label{lemma:ogd-contracting}
For a sequence of contracting convex decision sets $\K_1, \ldots \K_T$, $x_1 \in \K_1$ each with diameter at most $D$, a sequence of $G$-Lipschitz losses $\ell_1,\ldots, \ell_T$, and $\eta = \frac{D}{G\sqrt{T}}$, the regret of Algorithm \ref{alg:ogd-contracting} with respect to $\K_t$ is bounded by
\begin{align*}
\sum_{t=1}^T \ell_t(x_t) - \min_{x^* \in \K_T }\sum_{t=1}^T \ell_t(x^*) \leq&\;  GD \sqrt{T}.
\end{align*}
\end{lemma}

\begin{proof}
Let $x^* = \text{arg min}_{x \in \K_T} \sum_{t=1}^T \ell_t(x)$, and let $\nabla_t = \nabla \ell_t(x_t)$. First, note that
\begin{align*}
    \ell_t(x_t) - \ell_t(x^*) \leq&\; \nabla_t^{\top} (x_t - x^*)
\end{align*}
by convexity; we can then upper-bound each point's distance from $x^*$ by:
\begin{align*}
    \norm{x_{t+1} - x^*} =&\; \norm{ \Pi_{\K_{t+1}}(x_t - \eta \nabla \ell_t(x_t)) }  \leq \norm{x_t - \eta \nabla_t - x^*},
\end{align*}
using projection properties for convex bodies. Then we have
\begin{align*}
    \norm{x_{t+1} - x^*}^2 \leq&\; \norm{x_{t} - x^*}^2 + \eta^2 \norm{ \nabla_t }^2 - 2\eta \nabla_t^{\top} (x_t - x^*)
\end{align*} 
and 
\begin{align*}
    \nabla_t^{\top} (x_t - x^*) \leq&\; \frac{\norm{x_t - x^*}^2 - \norm{x_{t+1} - x^*}^2 }{2 \eta} + \frac{\eta \norm{\nabla_t}^2 }{2}.
\end{align*}
We can then conclude:
\begin{align*}
    \sum_{t=1}^T \ell_t(x_t) - \sum_{t=1}^T \ell_t(x^*) \leq&\; \sum_{t=1}^T \nabla_t^{\top} (x_t - x^*) \\ 
    \leq&\; \sum_{t=1}^T \frac{\norm{x_t - x^*}^2 - \norm{x_{t+1} - x^*}^2 }{2 \eta} + \frac{\eta  }{2} \sum_{t=1}^T \norm{\nabla_t}^2 \\ 
    \leq&\; \frac{\norm{x_T - x^*}^2}{2\eta} + \frac{\eta }{2} \sum_{t=1}^T \norm{\nabla_t}^2  \\
    \leq&\; \frac{D^2}{2\eta} + \frac{\eta }{2} \sum_{t=1}^T \norm{\nabla_t}^2 \\
    =&\; GD\sqrt{T} \quad \quad\quad \quad\quad \quad\quad \quad \quad\quad \quad\quad \quad (\text{when } \eta = \frac{D}{G\sqrt{T}})
\end{align*}
\end{proof} 

The bandit-to-full-information reduction is fairly standard as well, with a proof equivalent to that of e.g.\ Lemma 6.5 in \cite{ocobook}, modified for a full-information algorithm $\A$ for over contracting sets. 

\begin{lemma}
\label{lemma:btofi-contract}
Let $u$ be a fixed point in $\K_T$, let $\{ \ell_t : \K_t \rightarrow \R  | ~  t \in [T]\}$ be a sequence of differentiable loss functions, and let $\A$ be a first-order online algorithm that ensures a regret bound $\text{Regret}_{\K_T}(\A) \leq B_{\A}( \nabla \ell_1(x_1), \ldots, \nabla \ell_T(x_T))$ in the full-information setting for contracting sets $\K_1,\ldots, \K_T$. Define the points $\{x_t\}$ as $x_1 \leftarrow \A(\emptyset)$, $x_t \leftarrow \A(g_1, \ldots, g_{t-1})$, where $g_t$ is a random vector satisfying 
\begin{align*}
    \E[g_t | x_1, \ell_1, \ldots, x_{t}, \ell_{t}] =&\; \nabla \ell_t(x_t).
\end{align*}
Then for all $u \in \K_T$:
\begin{align}
    \E[\sum_{t=1}^T  \ell_t(x_t)] -  \sum_{t=1}^T  \ell_t(u) \leq&\; E[B_{\A}(g_1, \ldots, g_T)] \label{eqn:fi-regret}
\end{align}

\end{lemma}

\begin{proof}

Let $h_t : \K_t \rightarrow \R$ be given by:
\begin{align*}
    h_t(x) =&\; \ell_t(x) + \psi_t^{\top} x \text{, where } \psi_t = g_t - \nabla \ell_t(x_t).
\end{align*}
Note that $\nabla h_t(x_t) = g_t$, and so deterministically applying a first order algorithm $\A$ on $\{h_t\}$ is equivalent to applying $\A$ on stochastic first order approximations of $\{f_t\}$. Thus,
\begin{align*}
    \sum_{t=1}^T h_t(x_t)-\sum_{t=1}^T h_t(u) =&\; \leq B_{\A}(g_1 , \ldots, g_T).
\end{align*}
Using the fact that the expectation of each $\psi_t$ is 0 conditioned on history, and expanding, we get that
\begin{align*}
    \E[h_t(x_t)] =&\; \E[\ell_t(x_t)] + \E[\psi_t^{\top} x_t] \\ 
    =&\; \E[\ell_t(x_t)] + \E[\E[\psi_t^{\top} x_t | x_1,\ell_1,\ldots, x_{t}, \ell_{t} ] ] \\ 
    =&\; \E[\ell_t(x_t)] + \E[\E[\psi_t | x_1,\ell_1,\ldots, x_{t}, \ell_{t} ]^{\top} x_t ] \\ 
    =&\; \E[\ell_t(x_t)],
\end{align*}
and we can conclude by taking the expectation of Equation \ref{eqn:fi-regret} for any point $u \in \K_T$.
\end{proof}

The key remaining step is to observe that each $g_t$ is an unbiased estimator of $\nabla \hat{f}_t(x_t)$:
\begin{align*}
    \E[g_t | x_1, \hat{f}_1,\ldots, x_t, \hat{f}_t] =&\; \frac{n}{\delta} \E[ \phi_t u_t |  x_t, \hat{f}_t ] \\ 
    =&\; \frac{n}{\delta} \E[ \E[\phi_t | x_t, \hat{f}_t, u_t] \cdot u_t | x_t, \hat{f}_t ]  \\
    =&\; \E[ f_t(x_t + \delta u_t + \xi_t) u_t | x_t, \hat{f}_t  ]  \\
    =&\;  \E[ \hat{f}_t(x_t + \delta u_t) u_t  ]  \\
    =&\; \nabla \hat{f}_t(x_t),
\end{align*}
where the final line makes use the sphere sampling estimator for linear functions (as in e.g.\ Lemma 6.7 in \cite{ocobook}). This allows us to apply Lemma \ref{lemma:btofi-contract} to Algorithm \ref{alg:ogd-contracting}:
 
\begin{align*}  
  \sum_{t=1}^T \E[\phi_t] - \sum_{t=1}^T f_t(x^*) \leq&\; \sum_{t=1}^T \E[\hat{f}_t(x_t)] - \sum_{t=1}^T \hat{f}_t(x^*_{\delta,\epsilon}) + \frac{\delta TGD}{r} +  \frac{2 \epsilon TGD}{r} \\
  \leq&\; \text{Regret}_{COGD} \parens{ g_1, \ldots, g_T | \{\hat{f}_t\} } + \frac{\delta TGD}{r} +  \frac{2 \epsilon TGD}{r} \\
 \leq&\; \frac{D^2}{2\eta} + \frac{\eta }{2} \sum_{t=1}^T \norm{g_t}^2 + \frac{\delta TGD}{r} +  \frac{2 \epsilon TGD}{r}\\
  \leq&\; \frac{D^2}{2\eta} + \eta \frac{n^2}{2 \delta^2}T +  \frac{\delta TGD}{r} +  \frac{2 \epsilon TGD}{r} \quad \quad \quad \quad  (\text{def.\ of } g_t, \phi \leq 1) \\
   \leq&\; n GDT^{3/4} + \frac{ G D T^{3/4}}{r} +  \frac{2 \epsilon TGD}{r} \quad \quad   (\eta = \frac{D}{nT^{3/4}}, \delta = \frac{1}{T^{1/4}}). \\
\end{align*} 
\end{proof}

\section{Omitted Proofs for Section \ref{sec:uniform}}
\label{sec:main-alg-proof}
\subsection{Proof of Lemma \ref{lemma:unif-eird}}

\begin{proof}
Consider any memory vector $v \in \Delta(n)$. 
We can show constructively that there is some distribution of menus $z_U$ which induces the all-$\frac{1}{n}$ vector.

We construct $z_U$ in $\frac{1}{\tau} + 1$ stages for some $\tau > 0$, through a process where we continuously add weight $a_{z_j}$ to a sequence of distributions $\{ z_j | j \geq 1 \}$ over menus until the total weight $\sum_j a_{z_j}$ sums to 1. The uniform-inducing menu distribution $z_U$ will then be defined by taking the mixture of the menu distributions $z_j$ where each is weighted by $a_{z_j}$.

Consider the uniform distribution over all menus; continuously add weight to this distribution until some item (the one with the largest score in $M$) has selection weight $\tau/n$ (its selection probability under $M$ at memory vector $v$ in each distribution of menus $z_j$ considered thus far, weighted by $a_{z_j}$). While there are at least $k$ items with selection weight $\tau/n$, continuously add weight to the uniform distribution over all menus containing only items with weight below $\tau/n$.

Once there are fewer than $k$ items with selection weight at most $\tau/n$, we terminate stage 1. In general, for stage $i$, we always include every item with weight below $\tau i / n$ in the menu, with all others chosen uniformly at random.

Inductively, we can see that every item starts stage $i$ with at least weight $\tau(i-1)/n$ and at most $\tau i /n$, with at most $k-1$ items having weight less than $\tau(i-1)/n$.
Crucially, any item with weight less than $\tau i / n$ at the start of stage $i$ will reach weight $\tau i / n$ before any item starting at weight $\tau i / n$ reaches weight $\tau (i+1) / n$. Such an item is included in every menu until this occurs, resulting in a selection probability of at least $\frac{\lambda}{k}$ in each menu distribution considered, whereas any other item is only included in the menu with probability at most $\frac{k}{n}$, which bounds its selection probability in the menu distribution. As $\frac{\lambda}{k} \geq \frac{k}{n}$, the selection weight of items beginning stage $i$ below $\tau i / n$ reaches $\tau i / n$ no later than when the stage terminates. 

After stage $\frac{1}{\tau}$, every item has weight at most $\frac{1}{n}$ and at least $\frac{1}{n} - \frac{\tau}{n}$. We continue for one final stage until the sum of weights is 1, at which point every item has a final weight $p_{z_U} \in [\frac{1}{n} - \frac{\tau}{n}, \frac{1}{n} + \frac{\tau}{n}]$. Taking the limit of $\tau$ to zero gives us that $x_U$ is in $\texttt{IRD}(v, M)$ for any $v$, and hence $x_U$ is in $\texttt{EIRD}(M)$ as well.

Further, there is a distribution of menus $z_{b_i}$ where $i$ has probability $p_{b_i, i} = k / n$ and every other item $j$ has probability
\begin{align*}
    p_{b_i, j} =&\; \frac{1}{n} - \frac{k- 1}{n(n-1)}
\end{align*}

Here, we include $i$ in every menu and run the previous approach over the remaining $n - 1$ items for menus of size $k-1$, which we then augment with $i$. The required bound on $\lambda$ still holds for any $\lambda < 1$, as $\frac{k^2}{n} \geq \frac{(k-1)^2}{n-1}$ (for any $k\leq \sqrt{n}$, which holds as $\lambda < 1$). The selection probability of $i$ will be at least $\frac{\lambda}{k} \geq \frac{k}{n}$; we can take a mixture of this menu distribution with $z_U$ such that $p_{b_i, i} = \frac{k}{n}$ exactly. 

The convex hull of each $p_{b_i}$ is thus contained in $\texttt{EIRD}(M)$, as any point $p \in \convhull \{p_{b_i} | i \in [n]\}$ can be generated by taking the corresponding convex combination of menu distributions $z_{b_i}$. Any point $x \in \Delta(n)$ where $\norm{x_U - x}_{\infty} \leq \frac{k-1}{n(n-1)}$ can then be induced by taking mixtures of the $z_{b_i}$ menu distributions.

\end{proof}

\subsection{Subset-Uniform Distributions in \texttt{EIRD}}

\begin{lemma}\label{lemma:unif-subset-eird}
For any $\lambda$-dispersed $M$ where $\lambda \geq \frac{C k^2}{n}$, $\textup{\texttt{EIRD}}(M)$ contains the uniform distribution over any $\frac{n}{C}$ items. 

\end{lemma}
\begin{proof}
The proof of Lemma \ref{lemma:unif-eird} carries through directly for a universe with only $\frac{n}{C}$ items.
\end{proof}

\subsection{Implementing Near-Uniform Vectors}

\begin{lemma}\label{lemma:unif-eird-alg}
For any $\lambda$-dispersed $M$ where $\lambda \geq \frac{k^2}{n}$,
 for any point $x \in \Delta(N)$ satisfying 
\begin{align*}
    \norm{x - x_U}_{\infty} \leq&\; \frac{k-1}{n(n-1)},
\end{align*}
there is an adaptive strategy for selecting a sequence of menus over $t^*$ rounds, resulting in a $t^*$-round empirical distribution $\hat{x}$ such that $\norm{x - \hat{x}}_{\infty} \leq \gamma t^* + O(n)$ with probability at least $1 - 2n\exp(-\gamma^2 t^* /8)$, 
for any $\gamma$.
\end{lemma}

\begin{proof}
Our strategy will essentially correspond to the construction in Lemma \ref{lemma:unif-eird}, which shows that our vector is indeed in $\texttt{EIRD}(M)$.
For each item $i$, let $V_i = t^* \cdot x_i$ be the target number of rounds where $i$ is selected over the window. For any $t \leq t^*$ 
let $\hat{V}_{t, i}$ be the number of additional rounds an item must be selected before reaching its target, with $\hat{V}_{1, i} = V_i$. 
In each round $t$, construct a menu for the Agent by choosing the $k$ items with largest remaining counts $\hat{V}_{t, i}$, breaking ties uniformly at random, and decrement by 1 the count of the item selected in that round. Our approach will be to show that each item's final count under this process is close to its target in expectation after $t^*$ rounds, and use the sequence of expectations as rounds progress to define a martingale which will be close to its final expectation with high probability.

Let $\hat{V}_{t, \bot}$ denote the minimum value of $\hat{V}_{t, i}$ across items. Observe that our procedure maintains the invariant that $\hat{V}_{t, \bot}$ can only decrease in a round where at most $k-1$ items have remaining counts $\hat{V}_{t, i} > \hat{V}_{t, \bot}$. We will consider each round in which $\hat{V}_{t, \bot}$ decreases as the beginning of a ``trial'', and we will track the expectations of $\hat{V}_{t, i}$ over sequences of trials across two cases: 
\begin{itemize}
    \item Case 1: For every round $t$ at the start of a trial, we have had $\hat{V}_{t, i} - \hat{V}_{t, \bot} > 2$;
    \item Case 2: There has been some round $t$ at the start of a trial where $\hat{V}_{t, i } - \hat{V}_{t, \bot} \leq 2$.
\end{itemize}
When the first trial begins, we have at most $k-1$ items in Case 1, and items can never enter Case 1 after being in Case 2. We assume without loss of generality that we begin in a state where the first trial has just begun, as no prior rounds can increase the distance of any item from the minimum.

\emph{Case 1.}
Note that the probability of an item in the menu being selected in a given round is at least $\lambda/k \geq k/n$. 
We can upper-bound the expected distance of some count $\hat{V}_{t, i}$ from $\hat{V}_{t, \bot}$ by analyzing a ``pessimistic'' process where we assume that this minimum selection probability is tight, where every selection of an item other than item $i$ corresponds to the beginning of an ``event'', where the number of selections of $i$ in each event is geometrically distributed with parameter $p = 1 - \frac{k}{n}$. While these counts are not truly geometrically distributed, as the maximum number of selections is bounded, we will only need to analyze the probabilities of sums corresponding to items remaining in Case 1, in which case truncation does not affect the resulting distribution. Not every event corresponds to a new trial; there are deterministically at least $n-k$ events per trial, as every item begins a trial with a strictly higher count than $\hat{V}_{t, \bot}$, and so at least $n-k-1$ selections of items other than $i$ must occur before an item with minimum count can enter the menu (conditioned on $\hat{V}_{t, i}$ remaining above $\hat{V}_{t, \bot}$). 

Under this process, after $z$ events, the distribution of $\hat{V}_{t, i}$ is given by subtracting the sum of $z$ of the aforementioned geometric variables from $\hat{V}_{1, i}$, which is distributed according to a negative binomial:
\begin{align*}
    \Pr\brackets{\hat{V}_{1, i} - \hat{V}_{t, i} = y} =&\; { z + y - 1 \choose z - 1} \parens{\frac{k}{n} }^y \parens{1 - \frac{k}{n} }^z
\end{align*}
with mean $\frac{z(k/n)}{1 - k/n} = \frac{z k}{n - k} = \E[y]$ and variance $\frac{zk/n}{(1 - k/n^2)}$. After $z$ events, $\hat{V}_{1, \bot}$ has dropped by at most $\frac{z}{n-k}$. As such, by the time $\hat{V}_{t, \bot}$ reaches 0, we would also have that the expectation of $\hat{V}_{i, t}$ would reach 0 if we were to keep item $i$ in the menu at every round and allowed its count to drop below $\hat{V}_{t, \bot}$ without replacing it (and become negative); however, our process truncates (and enters Case 2) upon reaching 2 from the minimum, and so we can simply show that the contribution of the left tail of this distribution is small. 
Note that at the beginning of our process, we have $ \hat{V}_{1, i} - \hat{V}_{1, \bot} \leq \frac{t^*(k-1)}{n(n-1)}$, and so the expected difference from the minimum upon reaching $\hat{V}_{t, \bot} = 0$ while remaining in Case 1 is at most:
\begin{align*}
    \E[( \hat{V}_{t^*, i} - \hat{V}_{t^*, \bot})  \cdot  I(\text{Case 1}) ]  \leq&\; 2 + \sum_{y = 0}^{\hat{V}_{1, i} - 2} (2 + \hat{V}_{1, i} - y) { t^* - 1 \choose t^* - y - 1} \parens{\frac{k}{n} }^y \parens{1 - \frac{k}{n} }^{t^* - y} \\
    \leq&\; 2 +  \sum_{y = 0}^{\hat{V}_{1, i} - 2}  \frac{(2 + \hat{V}_{1, i} - y) t^*}{t^* - \hat{V}_{1, i}} { t^* \choose t^* - y} \parens{\frac{k}{n} }^y \parens{1 - \frac{k}{n} }^{t^* - y}. \\
\end{align*}
For any $y$ in this range we have 
going from $y-1$ to $y$:
\begin{align*}
  \frac{{ t^* \choose t^* - y} \parens{\frac{k}{n} }^{y} \parens{1 - \frac{k}{n} }^{t^* - y}}{{ t^* \choose t^* - y - 1} \parens{\frac{k}{n} }^{y-1} \parens{1 - \frac{k}{n} }^{t^* - y- 1}} =&\; \frac{t^* - y - 1}{y} \cdot \frac{ \frac{k}{n} }{1 - \frac{k}{n}} \\
   \geq&\; \frac{t^*}{\hat{V}_{1,i}} \cdot  \frac{k}{n} \\
    \geq&\; \frac{t^*}{1/n + k/n^2} \cdot  \frac{k}{n} \\
   \geq&\; \frac{k}{1 + k/n}
\end{align*}
which is greater than $1$ for any $k \geq 2$. As such, we can bound the tail summation by: 
\begin{align*}
    \E[( \hat{V}_{t^*, i} - \hat{V}_{t, \bot})  \cdot  I(\text{Case 1}) ]  \leq&\; 2 + \frac{ t^*}{t^* - \hat{V}_{1, i}} \cdot \sum_{y = 0}^{\infty} \parens{ \frac{1 + k/n}{k} }^y \\
    \leq&\; 2 + \frac{ t^*}{t^* - \hat{V}_{1, i}} \cdot \frac{{k}}{{k}  -1 - k/n } \\
    \leq&\; 5
\end{align*}
for $k \geq 2$ and sufficiently large $n$.

\emph{Case 2.}
Here we show that once an item has reached Case 2, its expected distance from $\hat{V}_{t, \bot}$ in any future round is at most a constant. 
Separating this analysis is necessitated by the fact that there exist edge cases where an item's expected distance from the minimum can be increasing (e.g. if all items start a trial at one above the minimum, an item can only have a decreasing distance if it becomes the next minimum, and can have a higher likelihood of remaining in the menu when the next trial begins).
Our approach will be to show by induction that, beginning from the first trial in Case 2, the distribution of item $i$'s distance from the minimum, where $p_y$ is the probability of distance $y$, satisfies: $$p_{y+1} \leq p_y / 2^{k/2 - 1}$$ 
for $y \geq 2$. This holds at the first trial in Case 2, as we have  $p_{y+1} = 0$ for each $y \geq 2$. An item can only have a distance increase of 1 in a given trial (if it is not picked in any of the at least $n-k$ rounds), which occurs with probability at most $\frac{1}{(1 + k/n)}^{n-k}\leq e^{-k/2} \leq \frac{1}{2^{k/2}}$, using that $n - k > n/2$ (which holds given that $k \geq 2$ and $n \geq k^2$). 
Further, using the same negative binomial process as in Case 1 to describe the number of selections of item $i$ in a given trial, we can see that $1/2$ upper bounds its density function after $n-k$ events for any valid setting of our parameters, and so the probability that an item is selected $j$ times, for $j$ such that it remains in every menu, is at most $1/2$. Letting $p^*$ describe the distribution after another trial, we can solve for:
\begin{align*}
    p^*_{y} =&\; p_{y-1} / 2^{k/2} + \sum_{j=0}^{\infty} p_{y+j} \cdot \Pr[{\text{drops } j+1 }] \\
    \leq&\; p_{y-1} / 2^{k/2} + p_y / 2; \\
    p_{y+1}^* =&\; z p_{y} / 2^{k/2 - 1}; \\
\end{align*}
using the induction hypothesis on $p$. As such, in any future trial, the expected distance from minimum can be given by:
\begin{align*}
    \E[ \hat{V}_{t^*, i} - \hat{V}_{t, \bot}  | \text{Case 2} ] \leq&\; 2 + \sum_{y=3}^{\infty} p_y \\
    \leq&\; 2 + \sum_{y=3}^{\infty} 2^{y(k/2 - 1)} \\
    \leq&\; 3.21
\end{align*}
for any $k \geq 3$. One can strengthen this to yield a constant sum for $k=2$ via a more delicate analysis on the upper bound of the negative binomial density function, which we omit.





\emph{Concentration Analysis.}
We now have that in either case, the expectation  $\E[( \hat{V}_{t^*, i} - \hat{V}_{t^*, \bot})  ]$ is a constant, for every item $i$. Given any current empirical counts
counts $\{\hat{V}_{t,i} : i \in [n]\}$ and scores for every item at any time $t$ (which we as the Recommender need not know), the distribution over subsequent items chosen is fully defined. Let $X_{t,i} = \Pr[~i \text{ chosen } | ~\{\hat{V}_{t-1,i} : i \in [n]\}, \{f_i(v_t)\} ]$.
For this process, we can now view each quantity $Y_{t,i} = (\hat{V}_{t-1,i} - \hat{V}_{t,i})$ as a Bernoulli random variable with mean $X_{t,i}$. Then we can define $Z_{t,i} = \sum_{h=1}^t Y_{h,i} - X_{h,i}$
as a martingale, where $\E[Z_{t,i}] = Z_{t-1, i}$ and $\abs{Z_{t,i} - Z_{t-1,i}} \leq 2$. Note that $\E[Z_{t^*,i}]$ is equal to $V_{i}$ up to a small constant $c_i$. We can then apply Azuma's inequality to get:
\begin{align*}
    Pr\brackets{\abs{ \hat{Z}_{t^*, i} - V_{i} - c_i} \geq \gamma t^* } \leq 2 \exp \parens{\frac{-\gamma^2 t^*}{8}}.
\end{align*}
These constants are independent of $t^*$, and will vanish when $t^*$ is sufficiently large.
\end{proof}

\subsection{Proof of Theorem \ref{thm:unif-regret}}

\begin{proof}

Let:
\begin{itemize}
    \item $F_{LL} = f_{LL}(\lambda, \alpha, n, \M)$ s.t. $\A_{\M}$ with $\beta / F_{LL}$ results in $\epsilon_{LL} = \frac{\epsilon \lambda k}{n}$;
    \item $F_Q = \frac{8 L \sqrt{n} k}{ \lambda} F_{LL}$;
    \item $t_{\text{query} } =  \frac{2 n}{k-1} \parens{\frac{F_{LL}}{\beta}}^2 \log\parens{ \frac{2n k S}{(k-1)\delta_{\text{query}}} } = \Tilde{\Theta}(1/\epsilon^2)$;
    \item $t_{\text{pad}} = \max \parens{  \frac{2 F_Q t_{\text{query} }}{\beta}, \frac{32 n^2 F_Q^2  \log(2/\delta_{\text{pad}})}{\beta^2}} = \Tilde{\Theta}(1/\epsilon^3)$;
    \item $t_{\text{move}} = \max \parens{  \frac{n(n-1) t_{\text{query}} }{k-1}, \frac{32 n^2 F_Q^2  \log(4S/\delta_{\text{move}})}{( 1 - 4k/n) \beta^2}, t_{\text{pad}} } = \Tilde{\Theta}(1/\epsilon^3)$;
    \item $t_0 = t_{\text{pad}} + S(2 \cdot t_{ \text{move}} + t_{\text{query}}) = \Tilde{\Theta}(1/\epsilon^3)$.
\end{itemize}

After running \texttt{UniformPad} via the first Lemma \ref{lemma:unif-eird-alg} construction for $t_{\text{pad}}$ steps, our empirical memory vector is within 
$\ell_{\infty}$ distance $\frac{\beta}{n F_Q}$ of 
$x_U$ with probability at least $1 - \delta_{\text{pad}}$.
We maintain the invariant that when calling $\texttt{MoveTo}(x)$ to reach some non-uniform vector $x$ from $x_U$, the $\ell_{\infty}$ distance between $x$ and $x_U$ is at most $\alpha$, and that after calling \texttt{Query}($x$) the current vector $x'$ (accounting for drift during sampling) has  $\ell_{\infty}$ distance at most $\alpha$ from $x_U$.

At any time $t < t_0$ when $\texttt{MoveTo}$ is called, the proportion of steps which the current invocation will contribute to the total history is at least:
\begin{align*}
    R_{\text{move}} =&\;  \frac{t_{\text{move}} }{t_{\text{pad}} + S(t_{\text{move}} + t_{\text{query}})} = O(1/S)
\end{align*}

Let $\alpha = \frac{k-1}{2{n}(n - 1)} \cdot R_{\text{move}}$ denote the radius of the $\ell_2$ ball around $x_U$
in which we permit queries for local learning. Any point $x$ within the $\alpha$-ball around the uniform vector can reach (or be reached from) the uniform vector with one call to $\texttt{MoveTo}(x)$, as their $\ell_{\infty}$ distance is at most $\alpha$, so some difference vector exists with mass $R_{\text{move}}$ and which satisfies the required norm bound.
For each input $x$, called from $x_t$, $\texttt{MoveTo}(x)$ applies the construction from Lemma \ref{lemma:unif-eird-alg} for the mass $t_{\text{move}}$ vector $y = x\cdot(t_{\text{move}}) -  x_t \cdot t$. This results in a total error of at most $\frac{\beta}{2 n F_Q} \cdot t_{\text{move}} + 1 \leq \frac{\beta}{n F_Q} \cdot t_{\text{move}}$ per item count with probability at least $1 - {\delta_{\text{move}}}$, as 
$$ t_{\text{move}}  \geq \frac{32 n^2 F_Q^2  \log(4S/\delta_{\text{move}})}{( 1 - 4k/n) \beta^2}. $$
This yields a total variation distance within $\frac{\beta}{2F_Q}$ for the entire memory vector when appended to the current history.

To run $\texttt{Query}(x)$, consider a set of $\frac{n}{k-1}$ menus, where item 1 appears in every menu and every other item appears in exactly one. Over the following $t_{\text{query} }$ rounds, play each menu $t_{\text{query} } \cdot \frac{k-1}{n}$ times and note the proportion of each item observed relative to item 1 when its menu was played. 
Each scoring function $f_i \in M$ is $L$-Lipschitz; we run $\texttt{Query}(x)$ for $t_{\text{query}}$ rounds, which can introduce a drift of at most $\beta / (2 F_Q)$ in total variation distance given the bound on $t_{\text{query}} $ in terms of $t_{\text{pad}}$. This drift results in a vector which remains within $\ell_{\infty}$ distance $2\alpha$ from $x_U$, and so $x_U$ can still be reached again in a single $\texttt{MoveTo}(x_U)$ call.

The empirical average memory vector over all menu queries (for any item) is within $\beta / F_Q$ total variation distance from $x$, and so the expected distribution of items differs from that at $x$ by at most 
$\beta / F_Q \cdot \frac{4 L\sqrt{n} k}{ \lambda} = \beta /(2 F_{LL})$ in $\ell_{\infty}$ distance. 
Each point's observed frequency differs from that expectation by at most $\beta /(2 F_{LL})$ with high probability.
For an item $i$ in the menu at a given round, we view whether or not it was chosen as a Bernoulli random variable, with mean equal to its relative score among items in the menu. Let $\bar{s}_{v, K,i}$ be the expected frequency of observing an item when the menu $K$ containing it is played, given the empirical sequence of memory vectors during those rounds $t_{\text{query} } \cdot \frac{k-1}{n}$, and let $\hat{s}_{v,K,i}$ be the true observed frequency. We then have:
\begin{align*}
    Pr\brackets{ \abs{ \bar{s}_{v, K,i} - \hat{s}_{v,K,i} } \geq \frac{\beta}{2 F_{LL}} } \leq&\; 2 e^{\parens{ -2  (\beta / 2F_{LL})^2  t_{\text{query}  } (k-1)/n }} \\
    =&\; 2 e^{\parens{ - (\beta / F_{LL})^2 t_{\text{query}}  (k-1) / (2n) }} \\
    \leq&\; \frac{\delta_{\text{query}} (k-1) }{ nk S},
    \end{align*}
    given that
    \begin{align*}
 t_{\text{query} } \geq&\;  \frac{2 n}{k-1} \parens{\frac{F_{LL}}{\beta}}^2 \log\parens{ \frac{2n k S}{(k-1)\delta_{\text{query}}} }.
\end{align*}
For item $1$ take the average over all menus, and rescale such that all scores sum to 1 (using the frequency of item $i$ relative to the frequency of item 1 when both were in the menu). Each score, and its error bound, will only shrink under the rescaling. 
This gives us score vector estimates $\hat{s}_x$ for each $x\in S$ with additive error at most $\frac{\beta}{F_{LL}}$ relative to the true frequency of item 1, and thus overall, where
$F_{LL} = f_{LL}(\lambda, \alpha, n, \M)$. This holds for every query simultaneously with probability $1 - \delta_{\text{query}}$. 

By the local learnability guarantee for $\M$, running  $\A_{\M}$ our results in a hypothesis $\hat{M}$ which has $\ell_2$ error at most $\epsilon_{LL} = \frac{\epsilon \lambda  k}{n}$ for any $x \in \Delta(n)$. 
In each round, the model and memory vector defines a space of feasible item distributions. 
This allows us to run \textsc{RC-FKM} for perturbations up to $\epsilon$.
We can represent each set $\texttt{IRD}(v_t, \hat{M})$ explicitly as the convex hull of normalized score estimates for every menu.

We implement \texttt{PlayDist}($x$) using current score estimates $\hat{M}(v_t)$ to generate a menu distribution which approximately induces the instantaneous item distribution $x$. Taking the convex hull over every menu's score vector under $\hat{M}$ yields a polytope representation of $\texttt{IRD}(v_t, \hat{M})$, which will contain our chosen action at each step.

\begin{lemma}\label{lemma:realizing}
Let $x$ be a point in $\textup{\texttt{IRD}}(v, M)$, and let $z \in \Delta({n \choose k})$ be a non-negative vector such that $\sum_{j \in {n \choose k}} z_j \cdot p_{K_j, v} = x$, where $K_j$ is the $j$th menu in lexicographic order. If the Recommender randomly selects a menu $K$ to show the Agent with probability according to $z$, then the Agent's item selection distribution is $x$.
\end{lemma}
\begin{proof}
The probability that the Agent selects item $i$ is obtained by first sampling a menu, then selecting an item proportionally to its score:
\begin{align*}
    \Pr[\textup{Agent selects } i] =&\;  \sum_{j \in {n \choose k}} z_j \cdot p_{K_j, v, i}  = x_i.
\end{align*}
\end{proof}

\begin{lemma} \label{lemma:play-dist}
Given $\hat{M}$ satisfying $\frac{e\lambda k}{n}$-accuracy and a target vector $x_t \in \textup{\texttt{IRD}}(v_t, \hat{M})$ generated by \textsc{RC-FKM}, there is a linear program for computing a menu distribution $z_t$ such that the induced item distribution $p_{z_t}$ satisfies
\begin{align*}
    \norm{ p_{z_t} - x_t }\leq&\; \epsilon.
\end{align*}
\end{lemma}
\begin{proof}
We can define a linear program to solve for $z$ with:
\begin{itemize}
    \item variables for $z_j \in [0,1]$, where $\sum_{j \in {n \choose k}} z_j = 1$,
    \item estimated induced distributions for each menu $\hat{p}_{K_j}$, and
    \item a constraint for each $i \in [n]$:
    \begin{align*}
        \sum_{j=1}^{ {n\choose k} } z_j \cdot \hat{p}_{K_j, i} =&\; x_{t, i}.
    \end{align*}
\end{itemize}
If $\norm{\hat{M}(x)/\hat{M}^* - M(x)/M^*_x} \leq \frac{\epsilon \lambda k}{n}$, then for any menu distribution $z$, we have that:
\begin{align*}
    \norm{p_{z, v} - \hat{p}_{z, v}  } \leq&\; \epsilon.
\end{align*}
Consider some menu $K$. The $\ell_2$ distance of score vectors restricted to the menu is at most $\frac{\epsilon \lambda k}{n}$, and each vector has mass at least $\frac{k \lambda}{n}$ by dispersion. Rescaling vectors to have mass 1 yields a bound of $\epsilon$, which is preserved under mixture (which is the induced distribution by Lemma \ref{lemma:realizing}), as well as when projecting into the $n-1$ dimensional space for \textsc{RC-FKM}, and so there is some perturbation vector $\xi_t$ with norm at most $\epsilon$ such that $z$ induces $x_t + \xi_t$.
\end{proof}

Note that the losses for \textsc{RC-FKM} can be $2G$-Lipschitz after the reparameterization where $x_{t,n} = 1 - \sum_{i=1}^{n-1} x_{t,i}$.
Any point satisfying within radius $r = \frac{k-1}{n(n-1)}$ from the uniform distribution in $n$ dimensions, feasible by Lemma \ref{lemma:unif-eird}, is within distance $r$ under the reparameterization as well, as we simply drop the term for $x_n$. 
The required radius surrounding $\mathbf{0}$ for \textsc{RC-FKM} of $r$ is thus satisfied, and we have that $\epsilon + \delta \leq r /T^{1/4} \leq r$. Further, the diameter of the simplex is bounded by $D = 2$. We can directly apply the regret bound of \textsc{RC-FKM} for these quantities, which holds with respect to $H_c \cap \texttt{EIRD}(\hat{M})$. By Lemma \ref{lemma:play-dist}, for any point $x \in \texttt{EIRD}(\hat{M})$, there is a point $x' \in \texttt{EIRD}({M})$ such that $\norm{x - x'} \leq \epsilon$. Projecting both points into $H_c$ cannot increase their distance by convexity, and so the optimality gap between the two sets is at most $\epsilon GT$.
Our total regret is at most the sum of:
\begin{itemize}
    \item Maximal regret for the learning runtime $G \cdot t_0$;
    \item The regret of \textsc{RC-FKM} over $T - t_0$ rounds;
    \item The gap between ${\texttt{EIRD}}(\hat{M})$ and ${\texttt{EIRD}}({M})$; and
    \item The union bound of each event's failure probability.
\end{itemize}
We can bound this by:
\begin{align*}
    \textup{Regret}_{C \cap \textup{\texttt{EIRD}}(M)}(T) \leq&\; G \cdot t_0 + 4n GT^{3/4} + \frac{4(\delta + 2\epsilon) G T}{r} + \epsilon GT+ (\delta_{\text{pad}} + \delta_{\text{move}} + \delta_{\text{query}})T \\
    =&\; \Tilde{O}(T^{3/4}) 
\end{align*}
when taking each of $\{ \delta_{\text{pad}}, \delta_{\text{move}}, \delta_{\text{query}} \} = \frac{1}{T^{1/4}}$. We can also bound the empirical distance from $H_c$.
\begin{lemma}
The diversity constraint is $O(\epsilon)$-satisfied by the empirical distribution $v_T$ with probability $1 - O(T^{-1/4})$.
\end{lemma}
\begin{proof}
Note that after $t_0$, the empirical distribution $v_{t_0}$ is within total variation distance $\frac{\beta}{2 F_Q} $ from $x_U$ (which is necessarily in $H_c$).
Further, each vector $x_t$ played by \texttt{RC-FKM} results in a per-round expected item distribution $y_t$ which lies in $H_c$ by the robustness guarantee. We can apply a similar martingale analysis as in Lemma \ref{lemma:unif-eird-alg} to the sequence of realizations of any item versus its cumulative expectation $\sum_{t > t^* } y_t$  to get a bound of (much less than) $\frac{\beta}{2 F_Q}$ in total variation distance as well, which is preserved under mixture. For any locally learnable class, $\beta = O(\epsilon)$. Note that for all the classes we consider, we have $\beta/(2 F_{Q}) \ll \epsilon$. Both events hold with probability $1 - O(T^{-1/4})$, as we can apply the same failure probabilities used for the learning stage for each. 

Note that for a constraint $H_c$ where $c$ is sufficiently bounded away from $\log(n)$ and for large enough $T$, this will in fact yield an empirical distribution which exactly satisfies $H_c$, as the weight $\tilde{O}(T^{3/4})$ uniform window will ``draw'' the empirical distribution back towards the center of $H_c$, as it dominates the total $\Tilde{O}(T^{1/2})$ total error bound (for the unnormalized empirical histogram $T \cdot v_T$) obtainable with a martingale analysis over the entire \textsc{RC-FKM} window.
\end{proof}

This completes the proof of the theorem.

\end{proof}